\documentclass[a4paper,12pt]{article}
\usepackage[margin=25mm]{geometry}
\usepackage{graphicx}
\usepackage{subcaption}
\usepackage{amsmath}
\usepackage{amsthm}
\usepackage{amsmath, amssymb}
\usepackage{epstopdf}
\usepackage{subcaption}
\usepackage{enumitem}
\usepackage{mathtools}
\usepackage{comment}
\usepackage{xcolor}
\usepackage[colorinlistoftodos]{todonotes}
\usepackage{adjustbox}
\usepackage{siunitx}
\usepackage[numbers,sort&compress]{natbib}
\title{Quiescence generates moving average in a stochastic epidemiological model with one host and two parasites}
\author{Usman Sanusi$^{1,2}$, Sona John$^{1,2}$, Johannes Mueller$^{2,3}$, Aur\'elien Tellier$^{1}$ \\
        \small $^{1}$Population Genetics, Technical University of Munich, 85354 Freising, Germany \\
        \small $^{2}$Department of Mathematics, Technical University of Munich, 85748 Garching, Germany  \\
        \small $^{2}$Institute for Computational Biology, Helmholtz Center Munich, 85764 Neuherberg, Germany  \\
}
\date{usman.sanusi@tum.de, sona.john@tum.de, johannes.mueller@mytum.de, aurelien.tellier@tum.de }
\begin{document}
\newpage

\maketitle
\textbf{Abstract}\\
Mathematical modelling of epidemiological and coevolutionary dynamics is widely being used to improve disease management strategies of infectious diseases. Many diseases present some form of intra-host quiescent stage, also known as covert infection, while others exhibit dormant stages in the environment. As quiescent/dormant stages can be resistant to drug, antibiotics, fungicide treatments, it is of practical relevance to study the influence of these two life-history traits on the coevolutionary dynamics. We develop first a deterministic coevolutionary model with two parasite types infecting one host type and study analytically the stability of the dynamical system. We specifically derive a stability condition for a five-by-five system of equations with quiescence. Second, we develop a stochastic version of the model to study the influence of quiescence on stochasticity of the system dynamics. We compute the steady state distribution of the parasite types which follows a multivariate normal distribution. Furthermore, we obtain numerical solutions for the covariance matrix of the system under symmetric and asymmetric quiescence rates between parasite types. When parasite strains are identical, quiescence increases the variance of the number of infected individuals at high transmission rate and vice versa when the transmission rate is low. However, when there is competition between parasite strains with different quiescent rates, quiescence generates a moving average behaviour which dampen off stochasticity and decreases the variance of the number of infected hosts. The strain with the highest rate of entering quiescence determines the strength of the moving average and the magnitude of reduction of stochasticity. Thus, it is worth investigating simple models of multi-strain parasite under quiescence/dormancy to improve disease management strategies.

\section{Introduction}
Dormancy or quiescence is a bet-hedging strategy common to many bacteria, fungi \cite{ref29,ref30}, invertebrates \cite{ref31}, and plants which evolves to dampen off the effect of bad conditions and maximize the reproductive output under good conditions \cite{ref32, ref10, ref33}. This bet-hedging in time occurs when the individual (bacteria, fungus, invertebrates) or the offspring of the individual (plants, invertebrates) enter dormancy with a low metabolic state for some period of time during which reproduction and evolution occurs in the active part of the population. The dormant individuals constitutes a reservoir, the so-called seed banks, and can re-enter the active population at a later time point. Dormancy (quiescence) evolves a bet-hedging strategy in response to unpredictable environments such as random variations of the abiotic conditions \cite{ref8}, competition under density-dependence regulation of the population \cite{ref49}, contact between a bacteria host and viruses \cite{ref50}, frequency- or density-dependent selection due to host-parasite coevolution \cite{ref42} or prey-predator interactions. Dormancy (quiescence) introduces overlap between generation and a storage effect which generates a time delay in the generation time \cite{ref1,ref15}. At the population level, dormancy is shown to slow down the rate of genetic drift, that is increasing the time to random loss or fixation of neutral alleles. Moreover, seed banks also slow down the action of natural selection by increasing the time to fixation (loss) of the positively (deleterious) selected alleles \cite{ref24,ref25,ref26}. We note the use of the term dormancy preferably for plant seeds or crustacean eggs (\textit{e.g. Daphnia sp.}), while quiescence refers to individual bacteria or fungi switching between "on" and "off" metabolic states \cite{ref51}. As we focus on microparasites in the following, we prefer the term quiescence from now on.  

Parasite quiescence is a strategy of microparasites (bacteria, fungi) becoming inactive inside an infected host for some period of time. During this period, the disease does not progress in the host and the host can express symptoms or be asymptomatic. Parasite quiescence has well known but yet underappreciated consequences for disease management. During quiescence, the parasite are often resistant to the application of drugs, antibiotics or fungicides \cite{ref34,ref35,ref36,ref37}. Furthermore, applying antibiotics can trigger the switching of bacteria from active to the inactive (quiescent) state. \textit{Plasmodium falciparum}, the main agent of malaria, has the ability to lurk in the hepatocytes of some patients, remaining inactive but being resistance to drug treatments, causing later on disease relapse \cite{ref27,ref37,ref42}. \textit{P. vivax}, another malarial agent, exhibits also the ability to become dormant in the liver of a host for some weeks, months even up to a year or more, which makes the task to eradicate the disease difficult \cite{ref45,ref46,ref47}. Therefore, it is important to determine the 1) conditions for the evolution of parasite quiescence, and 2) influence of quiescence on the sustainability of parasite populations.
A key theoretical study on the evolution of quiescence in animal parasites \cite{ref9} shows that covert infection is not likely to be a parasite Evolutionary Stable Strategy (ESS) in an epidemiological model with one host and one parasite genotype. Parasite quiescence would only evolve if there were substantial fluctuations in the host population size or seasonal variations in transmission rates. Therefore, the authors state that their “models predict low rates of covert infection, which does not reflect the consistent high levels that are found in some host–parasite systems”. Based on a modelling framework with fixed population sizes but two hosts and two parasite types, the host population can evolve dormancy as an ESS as a result of the parasite pressure and coevolutionary dynamics \cite{ref10}. While more theoretical work is needed to decipher the conditions for the evolution of parasite quiescence/dormancy, likely involving a combination of temporally variable environmental and coevolutionary pressures, we focus in the present study on the consequence of quiescence for the stability and outcome of host-parasite coevolutionary dynamics. As a first step in this direction, we consider here a model with one host and two parasite strains (or types). 

Indeed, one host population under pressure by several parasite strains, or even several parasite species, is the rule rather than the exception \cite{ref6,ref7}. Considering the epidemiological dynamics under competition/co-infection between strains is important \cite{ref8} to predict the evolution of parasite virulence, that is disease induced death rate of host \cite{ref41}. We are interested here in understanding the epidemiological dynamics of a single host type infected by one of the two parasite strains exhibiting quiescence. We ask whether quiescence affects the parameters for which two strains can co-exist or competitively exclude one another. Furthermore, the maintenance of several strains, the persistence of disease as endemic or the persistence of the host population are affected by stochastic processes. Disease epidemics are subjected to stochasticity at various levels, the main one being in the transmission rate, and thus stochastic approaches are required to predict the outcome of epidemics. While the deterministic model of epidemiology successfully captures the behaviour when the size of host and parasite populations are large, stochasticity can affect the outcome of the dynamics for small sizes significantly \cite{ref11,ref12,ref13,ref14}. Quiescence affects the size of the parasite active population and thus possibly the epidemiological dynamics. We hereby hypothesize that quiescence may also affects the outcome of stochasticity on the co-existence of our two parasite strains epidemiological model. 

In the first part we describe our epidemiological model with changes in the number of healthy and infected host individuals over time under quiescence of both parasite strains. We then derive a stability condition for the dynamical ODE system. In the second part of the study, we introduce stochasticity in disease transmission and derive a Fokker-Planck equation of the Continuous Time Markov Chain model. Lastly, we perform some numerical study on the model behaviour under stochasticity. We show that for symmetric case i.e when the infected class are identical and quiescence phases are also identical, quiescence increases the variance, and decrease it when the rate of infection is small. For asymmetric case i.e when the infected class as well as the quiescence phases are  not identical, quiescence has a major effect in reducing the intensity of the noise in the stochastic process, whenever the rate of entering (or exiting) quiescence differ between strains. By analogy, we term this phenomenon as \textit{moving average}.

\section{Deterministic model with quiescence} 
\subsection{Model description}
\label{DeModel}
Our model is similar in essence to classic epidemiological models \cite{ref1,ref2,ref8,ref20,ref22,ref28}. Here we consider one host population and two  parasite strains, thus the population is divided into five mutually exclusive compartments: one healthy susceptible host compartment $H$, two infected host, $I_1$ and $I_2$, infected by parasite of type $1$ and $2$ respectively, and two quiescence compartments $Q_{1}$ and $Q_{2}$, comprise the infected individuals $I_{1}$ and $I_{2}$ for which the parasite is in the quiescent state. We define the following system of ordinary differential equations describing the rate of change of the number of individuals in each compartment. 
\begin{equation}
\begin{aligned}
& \frac{dI_1}{dt} = \beta_1 H I_1-\rho_1I_1-d I_1-\gamma_1  I_1 -\nu_1I_1+\zeta_1Q_1+\epsilon_1\\
& \frac{dI_2}{dt} =  \beta_2 H I_2-\rho_2I_2-d I_2-\gamma_2 I_2-\nu_2I_2+\zeta_2Q_2 +\epsilon_2\\
& \frac{dH}{dt} =  \Lambda- \beta_1H  I_1- \beta_2H I_2 -dH +\nu_1I_1+\nu_2I_2\\
& \frac{dQ_1}{dt} = \rho_1I_1-\zeta_1Q_1-dQ_1 \\
& \frac{dQ_2}{dt} = \rho_2I_2-\zeta_2Q_2-dQ_2
\end{aligned}
\label{Model1}
\end{equation}

Where  $\Lambda$  is the constant birth rate of healthy host and $d$ to is the natural death rate, $\gamma_1$ and $\gamma_2$ are the disease induced death rate or (virulence) caused by parasite $1$, and $2$ respectively. Similarly all other parasite specific parameters such as disease transmission rate $\beta$, recovery rate $\nu$, rate at which parasite switches to quiescence $\rho$ and the switching back rate $\zeta$ are defined for each parasite strains separately. 
The parameters $\epsilon_1$ and $\epsilon_2$ are the rates of incoming migration of parasite $1$ and $2$ respectively from an outside compartment/population. We assume 1) there is no free living parasite, i.e the parasite cannot live outside its host, 2) the absence of multiple infection so that strains $1$ and $2$ of the parasite are mutually exclusive on one host, and 3) no latency period for the parasite, hence, the infected persons are infectious immediately after infection.  Note that the model reduces to a simple model of one susceptible host and two infected host types ($SI_1I_2S$, referred to as system without quiescence) when setting the quiescence parameters equal to zero (Appendix C). In the present study we are particularly interested in following the number of hosts infected by parasite $1$ or $2$ and to study conditions for which both types of parasites are maintained. We therefore assume constant birth rate, to ensure a non-explosive process when moving to the stochastic version of our model. We finally introduce the parameters  $\epsilon_1$ and $\epsilon_2$ to promote the coexistence of both strains at the equilibrium and to guarantee a unique steady state solution in the continuous time Markov chain version of the model (see below, Stochastic model)  

\subsection{Steady state solutions}
In this section we find the steady state/equilibrium solutions of the system. First, we analyse the system without inflow of new infection to the population ($\epsilon_1 =\epsilon_2 =0$). This simple system generically has the three equilibrium states: 1) a disease free equilibrium in which both parasite strains die off and are removed from the system (yielding $I_1=I_2=Q_1 = Q_2= 0$), 2) two-boundary equilibria at which a single parasite strain survive \textit{i.e.} competitive exclusion when parameters of the model are non-symmetric (yielding in either $I_1=Q_1 = 0$ or $I_2=Q_2=0$). In the non-generic case that we have symmetric parameters, we have line of stationary solutions. By evaluating the Jacobian matrix of the system, one can evaluate the stability conditions for these equilibria. To ensure the existence of unique polymorphic equilibrium, we introduce two parameters for invasion/immigration rates namely, $\epsilon_{1}$ and $\epsilon_{2}$ which are greater than zero. The introduction of these two parameters results in moving the disease free as well as one of the boundary equilibria to the negative cone \textit{i.e.} makes them to have negative values which is biologically meaningless. We are thereafter left with only one polymorphic equilibrium which is biologically meaningful. Henceforth, we focus on the analysis of the polymorphic equilibrium for which both parasite strains are maintained in the system. We show the existence and uniqueness of this endemic equilibrium under mild conditions (for more details, see Appendix A).

\subsection{Stability analysis}
An $n \times n$ Jacobian matrix $P$ is said to be stable, and thus an equilibrium being locally stable, if all its eigenvalues lie on the left half plane. As it may be impractical to determine the stability of matrix analytically \cite{ref1}, we use here the Lyapunov theorem to determine if the system is stable. We apply the Routh-Hurwitz criteria \cite{ref1,ref39,ref40}, which can be also cumbersome if the matrix is of high dimension. In this section we therefore derive the stability condition for a generic $5 \times 5$ matrix $G$ with parasite quiescence by reducing our system to $3 \times 3$ which is more easily amenable to computation.

The Jacobian of system in equation (\ref{Model1}) evaluated at equilibrium is given as follows 
\begin{equation*}
 G = 
 \begin{pmatrix}
  \beta_1H^*-\rho_1-\gamma_1 -\nu_1-d&0 &  \beta_1 I_1^* &  \zeta_1 & 0\\ 
  
 0& \beta_2H^*-\rho_2-\gamma_2 -\nu_2-d &  \beta_2 I_2^* &  0  & \zeta_2\\ 

  - \beta_1H^*+\nu_1 &  - \beta_2H^*+\nu_2 &  -\beta_1 I_1^*-\beta_2 I_2^*-d & 0     & 0\\ 
     \rho_1 & 0&  0 & - \zeta_1-d&0 \\ 
   0&   \rho_2 & 0&  0 & - \zeta_2-d
 \end{pmatrix}.
 \end{equation*}
 
 If we define a matrix 
 \begin{equation}
 A\in ((a_{i,j})) \in \mathbb{R}^{3\times 3}
 \label{hurwitz}
 \end{equation}
to be the Jacobian matrix evaluated at equilibrium of the system without quiescent described in appendix C. We introduce $B=G+dI$, such that the spectrum of $B$ is just the shifted spectrum of $G$. Indeed, the stability of $B$ implies stability of $G$.\\
Let 
\begin{equation}
 B = 
 \begin{pmatrix}
  a_{11}-\rho_1 & a_{12} &  a_{13} &  \zeta_1 & 0\\ 
  a_{21} & a_{22} -\rho_2&  a_{23}&  0  & \zeta_2\\ 
a_{31} &  a_{32} &  a_{33}  & 0     & 0\\ 
     \rho_1 & 0&  0 & - \zeta_1&0 \\
   0 &   \rho_2 & 0&  0 & - \zeta_2
 \end{pmatrix}.
\label{matrixB}
 \end{equation}

\newtheorem{mypro1}{Proposition}
 \begin{mypro1} 
Let  $ 3 \times 3$ matrix $A$  be a Jacobian  matrix of system without quiescence phase and we also define 
\begin{equation}
 \begin{aligned}
 & a_1 = -\text{tr(A)} = -a_{11}-a_{22}-a_{33}, \\
 & a_2 = a_{11} a_{22}+a_{11} a_{33} +a_{22} a_{33}-a_{23} a_{32}-a_{12} a_{21}-a_{13} a_{31} ,\\
 & a_3 = -\text{det(A)}.
 \end{aligned}
 \end{equation}
The matrix $A$ in \ref{hurwitz} is stable if and only if 
  \begin{equation}
  \text{tr(A)} < 0,\quad  \text{det (A)} < 0 \quad \text{and}\quad  a_2 >  0 .
 \end{equation}
  \end{mypro1}
The above proposition 1 is simply a reformulation of Routh-Hurwitz criteria, for more details (see \cite{ref1,ref39,ref40}). We now find a criteria for stability of $B$ under the following proposition.
\begin{mypro1}
The following three statements are equivalent on matrix $B$ above: \\
\textit{Statement 1}:\\ 
 The matrix $B$ in \ref{matrixB} is stable for all $\rho_1, \rho_2, \zeta_1, \zeta_2 > 0 .$ \\
 \textit{Statement 2 :}\\
$  b_1 > 0,\quad  b_2 > 0,\quad  b_3 > 0, \quad b_4 > 0, \quad b_5 > 0, \quad  b_1 b_2 b_3 > b_3^2+b_1^2 b_4, $\\ 
 $ (b_1 b_4-b_5)(b_1 b_2 b_3-b_3^2-b_1^2 b_4) > b_5 (b_1 b_2-b_3)^2+b_1 b_5^2 $ \\ 
  for all $\rho_1, \rho_2, \zeta_1, \zeta_2 > 0.$ \\
\textit{Statement 3:} \\
$\text{det(A)} < 0,\quad  \text{tr(A)} \le 0,\quad a_2 > 0, \quad a_{11} \le 0, \quad a_{22} \le 0, a_{33}\le 0,$\\
$  a_{13} a_{31} \le a_{11} a_{33} , \quad  a_{23} a_{32} \le a_{22} a_{33}. $ 
\end{mypro1}

The above statements are technically equivalent in the sense that for the system in (\ref{Model1}) to be stable it must satisfy one of the given statements. Whenever the second statement is satisfied, the third statement is also automatically satisfied. This proposition is a generalisation of the theorem in \cite{ref1} and we use the same method as he does (see Appendix B for the proof of the proposition 2 above, we did prove the stability of a generic matrix $B$ as defined in \ref{matrixB}). The conditions in \textit{statement 3} of the above proposition can be used to prove that the endemic equilibrium of (\ref{Model1}) is locally asymptotically stable.  
\section{Stochastic Analysis}
\subsection{Transition probabilities}
This section defines a stochastic version to the deterministic model as described in  equation (\ref{Model1}) of section \ref{DeModel}. We add stochasticity occurring at any of the possible transition of individuals between classes (birth and death). The transition probabilities of jumping from one state (e.g. infected quiescent) to the another state (e.g. infected) are defined bellow. We choose $\Delta t$ very small so that during this time interval only one event occurs. The proportion of healthy population is $H $, the proportion of infected by parasite 1 population is $I_1$, the proportion of infected by parasite 2 population is $I_2 $, the proportion of population in quiescence compartment infected by parasite 1 is $Q_1$ and the proportion of population in quiescence compartment infected by parasite 2 is $Q_2.$ The possible changes are either $H+1, H-1, I_1+1, I_1-1, I_2+1, I_2-1, Q_1+1, Q_1-1, Q_2+1,Q_2-1$ or no change at all. Therefore, our stochastic process is a birth and death process. The one step transition probabilities are given in table \ref{quiescIItrans1}: 

\begin{table}[h!]
\begin{adjustbox}{width=0.95\textwidth,center}
\begin{tabular}{lll}
\hline
Type & Transition & Rate\\
\hline
birth of healthy host $H$ 
& $(H_t,{I_1}_t,{I_2}_t, {Q_1}_t, {Q_2}_t)\rightarrow({H_t}+1,{I_1}_t,{I_2}_t, {Q_1}_t, {Q_2}_t)$
&$\Lambda  \Delta t  +\tiny {o}\Delta(t)$\\
natural death of $H$
& $(H_t,{I_1}_t,{I_2}_t, {Q_1}_t, {Q_2}_t)\rightarrow({H_t}-1,{I_1}_t,{I_2}_t, {Q_1}_t, {Q_2}_t)$
&$ d H \Delta t  +\tiny {o}\Delta(t) $\\
infection of $H$ by $I_1$
& $(H_t,{I_1}_t,{I_2}_t, {Q_1}_t, {Q_2}_t)\rightarrow({H_t}-1,{I_1}_t+1,{I_2}_t, {Q_1}_t, {Q_2}_t)$
&$\beta_1 H I_1  \Delta t +\tiny {o}\Delta(t)$\\
infection of $H$ by $I_2$
& $(H_t,{I_1}_t,{I_2}_t, {Q_1}_t, {Q_2}_t)\rightarrow({H_t}-1,{I_1}_t,{I_2}_t+1, {Q_1}_t, {Q_2}_t)$
&$\beta_2 H I_2  \Delta t +\tiny {o}\Delta(t)$\\
death of $I_1$
&$(H_t,{I_1}_t,{I_2}_t, {Q_1}_t, {Q_2}_t)\rightarrow({H_t},{I_1}_t-1,{I_2}_t, {Q_1}_t, {Q_2}_t)$
&$(d+\gamma_1)I_1  \Delta t+\tiny {o} \Delta(t)$\\
death of $I_2$
&$(H_t,{I_1}_t,{I_2}_t, {Q_1}_t, {Q_2}_t)\rightarrow({H_t},{I_1}_t,{I_2}_t-1, {Q_1}_t, {Q_2}_t)$
&$(d+\gamma_1)I_2  \Delta t+\tiny {o} \Delta(t)$\\
recovery  $I_1$  \& replacement with $H$
& $(H_t,{I_1}_t,{I_2}_t, {Q_1}_t, {Q_2}_t)\rightarrow({H_t}+1,{I_1}_t-1,{I_2}_t, {Q_1}_t, {Q_2}_t)$
&$\nu_1I_1 \Delta t +\tiny {o}\Delta(t)$\\
recovery $I_2$ \& replacement with $H$
& $(H_t,{I_1}_t,{I_2}_t, {Q_1}_t, {Q_2}_t)\rightarrow({H_t}+1,{I_1}_t1,{I_2}_t-1, {Q_1}_t, {Q_2}_t)$
&$\nu_2I_2 \Delta t +\tiny {o}\Delta(t)$\\
immigration to $I_1$
&$(H_t,{I_1}_t,{I_2}_t, {Q_1}_t, {Q_2}_t)\rightarrow({H_t},{I_1}_t+1,{I_2}_t, {Q_1}_t, {Q_2}_t)$
&$\epsilon_1   \Delta t  +\tiny {o}\Delta(t) $\\
immigration to $I_2$
&$(H_t,{I_1}_t,{I_2}_t, {Q_1}_t, {Q_2}_t)\rightarrow({H_t},{I_1}_t,{I_2}_t+1, {Q_1}_t, {Q_2}_t)$
&$\epsilon_2   \Delta t  +\tiny {o}\Delta(t) $\\
go quiescent $I_1$ \& birth of $Q_1$
& $(H_t,{I_1}_t,{I_2}_t, {Q_1}_t, {Q_2}_t)\rightarrow({H_t},{I_1}_t-1,{I_2}_t, {Q_1}_t+1, {Q_2}_t)$
&$\rho_1I_1 \Delta t +\tiny {o}\Delta(t)$\\
go quiescent $I_1$ \& birth of $Q_1$
& $(H_t,{I_1}_t,{I_2}_t, {Q_1}_t, {Q_2}_t)\rightarrow({H_t},{I_1}_t,{I_2}_t-1, {Q_1}_t, {Q_2}_t+1)$
&$\rho_2I_2 \Delta t +\tiny {o}\Delta(t)$\\
wake-up $Q_1$ \& replacement with $I_1$
& $(H_t,{I_1}_t,{I_2}_t, {Q_1}_t, {Q_2}_t)\rightarrow({H_t},{I_1}_t+1,{I_2}_t, {Q_1}_t-1, {Q_2}_t)$
&$\zeta_1Q_1 \Delta t +\tiny {o}\Delta(t)$\\
wake-up $Q_2$ \& replacement with $I_2$
& $(H_t,{I_1}_t,{I_2}_t, {Q_1}_t, {Q_2}_t)\rightarrow({H_t},{I_1}_t,{I_2}_t+1, {Q_1}_t, {Q_2}_t-1)$
&$\zeta_2Q_2 \Delta t +\tiny {o}\Delta(t)$\\
natural death of $Q_1$
& $(H_t,{I_1}_t,{I_2}_t, {Q_1}_t, {Q_2}_t)\rightarrow({H_t},{I_1}_t,{I_2}_t, {Q_1}_t-1, {Q_2}_t)$
&$ dQ_1 \Delta t +\tiny {o}\Delta(t)$ \\
natural death of $Q_2$
& $(H_t,{I_1}_t,{I_2}_t, {Q_1}_t, {Q_2}_t)\rightarrow({H_t},{I_1}_t,{I_2}_t, {Q_1}_t, {Q_2}_t-1)$
&$ dQ_2 \Delta t +\tiny {o}\Delta(t)$
\end{tabular}
\end{adjustbox}
\caption{Transitions for the quiescence model 1.}\label{quiescIItrans1}
\end{table}

\subsection{Stochastic Simulations}
\begin{figure}[h!]
\includegraphics[width= 15 cm]{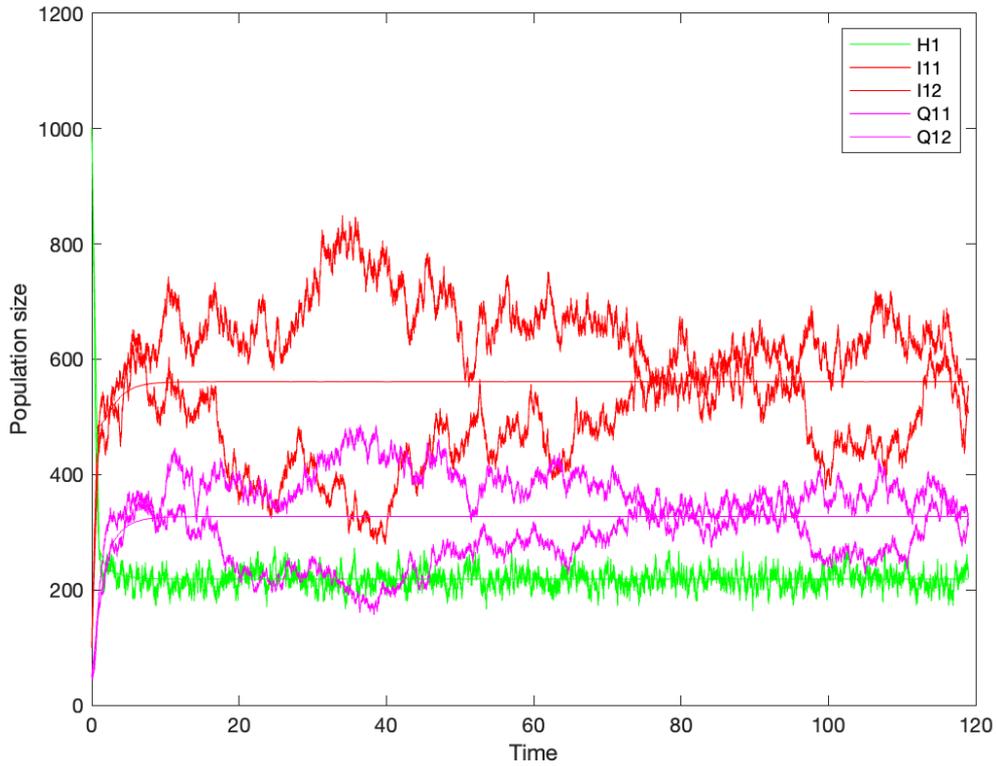}
\caption{Numerical simulation of the deterministic model compared with stochastic simulation using Gillespie's algorithm; initial population size is $H =1000,$ $I_1=100, I_2=100, Q_1 = Q_2 = 50.$ The values of the parameters are $ \beta_1 = \beta_2= 0.005, \Lambda = 1000,  d = 0.5,\nu_1= \nu_2 = 0.3, \gamma_1 =  \gamma_2 = 0.003, \epsilon_1 = \epsilon_2 = 0.6, \zeta_1 = \zeta_2 = 0.7, \rho_1 = \rho_2 = 0.7$. The stochastic realisations fluctuate about the equilibrium of the deterministic trajectories.}
\label{combi}
\end{figure}

In order to test the validity of our assumptions to analyse the stochastic system, we used Gillespie's algorithm (see \cite{ref17,ref18,ref21}) to generate stochastic realisations/sample paths of the birth and death processes (Figure \ref{combi}). In (Figure \ref{combi}), the stochastic trajectories fluctuates around the deterministic equilibrium as predict by equation (\ref{Model1}). Please note that there are only three (3) curves in the deterministic trajectories while there five (5) in the stochastic realisation. This is to due the fact that we chose symmetric parameter values of the model, so $I_1 =I_2$ and $Q_1=Q_2$ in the deterministic setting, but not in the stochastic version.

\section{Master Equation}
 The forward Kolmogorov differential equations also known as Mater Equation which describes the rate of change of these probabilities is given in table \ref{quiescIItrans1}. The master equation describes the evolution of the disease individuals at the early times of the infection. To understand the long term dynamics, we need to derive its corresponding Fokker-Planck equation. \\ 
 Let  $ p(i,j,k,l,m)(t)  =  \text{Prob} \{H(t) = i, I_{1}(t)= j, I_{2}(t) = k, Q_1(t) = l, Q_2(t) = m  \}, $ then

 \begin{equation}
\begin{aligned}
 \frac{\mathrm dp_{(i,j,k,l,m)}}{\mathrm d t}  =  & \Lambda p_{(i-1,j,k,l,m)}+ d (i+1)p_{(i+1,j,k,l,m)} +\beta_1( i+1) (j-1)p_{(i+1,j-1,k,l,m)} \\
 & +(d  +\gamma_1)(j+1)p_{(i,j+1,k,l,m)}+\beta_2 (i+1)(k-1)p_{(i+1,j,k-1,l,m)} \\
 &+(d+\gamma_2) (k+1)p_{(i,j,k+1,l,m)} +\nu_1 (j+1)p_{(i-1,j+1,k,l,m)}+\nu_2 (k+1)p_{(i-1,j,k+1,l,m)}\\
 & + \epsilon_1 p_{(i,j-1,k,l,m)}+ \epsilon_2 p_{(i,j,k-1,l,m)}+\rho_1(j+1) p_{(i,j+1,k,l-1,m)}+\rho_2(k+1) p_{(i,j,k+1,l,m-1)} \\
 &+\zeta_1(l+1) p_{(i,j-1,k,l+1,m)}+\zeta_2(m+1)p_{(i,j,k-1,l,m+1)} \\ 
 &+d(l+1)p_{(i,j,k,l+1,m)} +d(m+1)p_{(i,j,k,l,m+1)} \\
& -\Big[\Lambda+ di+\beta_1 ij 
+( d+\gamma_1)j+\beta_2 ik+(d+\gamma_2) k+\nu_1 j+\nu_2 k \\
&+ \epsilon_1 + \epsilon_2 +\rho_1j+\rho_2 k +\zeta_1 l+\zeta_2 m +d l+d m\Big] p_{(i,j,k,l,m)}
\end{aligned}
\label{Kolmo}
\end{equation}
This master equation (\ref{Kolmo}) is then used to work out \textit{Kramers-Moyal expansion} that led to the derivation of the \textit{Fokker-Planck equation} below.

\subsection{Fokker-Planck equation of the model}
To understand the long term dynamics of the master equation (\ref{Kolmo}), we need to derive the corresponding Fokker-Planck equation. The Fokker-Planck equation describes further the rate of change of transitions probabilities described in table \ref{quiescIItrans1}. We can also find the long term distribution of variables. \\ \\

Now, let 

$$p(i,j,k,l,m) =  \int_{ih-\frac{h}{2}}^{ih+\frac{h}{2}}\int_{jh-\frac{h}{2}}^{jh+\frac{h}{2}}\int_{kh-\frac{h}{2}}^{kh+\frac{h}{2}} \int_{lh-\frac{h}{2}}^{lh+\frac{h}{2}} \int_{mh-\frac{h}{2}}^{mh+\frac{h}{2}}u(x_1,x_2,x_3,x_4,x_5)dx_1dx_2dx_3dx_4 dx_5 +o(h^6),$$
let also $x_1 = ih,x_2 = jh, x_3 = kh, x_4 = lh, x_5 = mh$ and $h = \frac{1}{N}$. We then performed \textit{Kramers-Moyal expansion} to derived the following \text{Fokker-Planck equation} which is given as follows. 
\begin{equation}
\begin{aligned}
\partial_t u(x_1, \dots, x_5,t) = -& \partial_{x_1} \{h \lambda-dx_1-\beta_1 x_1x_2-\beta_2 x_1x_3+\nu_1 x_2+\nu_2 x_3\} u(x_1, \dots, x_5,t) \\
- & \partial_{x_2} \{\beta_1 x_1x_2-(d+\gamma_1)x_2-\nu_1 x_2-\rho_1 x_2+\zeta_1 x_4+\epsilon_1\} u(x_1, \dots, x_5,t) \\
 - & \partial_{x_3} \{\beta_2 x_1x_3-(d+\gamma_2)x_2-\nu_2 x_2-\rho_2 x_3+\zeta_2 x_5+\epsilon_2\} u(x_1, \dots, x_5,t) \\
 - & \partial_{x_4} \{\rho_1 x_2-\zeta_1 x_4-dx_4\} u(x_1, \dots, x_5,t) \\
  - & \partial_{x_5} \{\rho_2 x_3-\zeta_2 x_5-dx_5\} u(x_1, \dots, x_5,t) \\
+ & \frac{h}{2} \partial_{x_1x_1} \{h \lambda+dx_1+\beta_1 x_1x_2+\beta_2 x_1x_3+\nu_1 x_2+\nu_2 x_3\} u(x_1, \dots, x_5,t)  \\
+ & \frac{h}{2} \partial_{x_2 x_2} \{\beta_1 x_1x_2+(d+\gamma_1)x_2+\nu_1 x_2+\rho_1 x_2+ h \epsilon_1\} u(x_1, \dots, x_5,t) \\
+  & \frac{h}{2} \partial_{x_3 x_3} \{\beta_2 x_1x_3+(d+\gamma_2)x_3+\nu_2 x_3+\rho_2 x_3+ h \epsilon_2\} u(x_1, \dots, x_5,t) \\
+ &\frac{h}{2}  \partial_{x_4x_4} \{\rho_1 x_2+\zeta_1 x_4+dx_4\} u(x_1, \dots, x_5,t) \\
+  &\frac{h}{2}  \partial_{x_5x_5} \{\rho_2 x_3+\zeta_2 x_5+dx_5\} u(x_1, \dots, x_5,t) \\
-&h \partial_{x_1x_2} \{\beta_1 x_1x_2+\nu_1 x_2\} u(x_1, \dots, x_5,t) \\
-&h \partial_{x_1x_3} \{\beta_2 x_1x_3+\nu_1 x_3\} u(x_1, \dots, x_5,t) \\
 -&h  \partial_{x_2x_4} \{\rho_1 x_2+\zeta_1 x_4\} u(x_1, \dots, x_5,t) \\
- &h  \partial_{x_3 x_5} \{\rho_2 x_3+\zeta_2 x_5\} u(x_1, \dots, x_5,t) 
\end{aligned}
\label{FPeq}
\end{equation}

\subsection{Linear Transformation of the Fokker-Planck equation}
\begin{figure}[h!]
\includegraphics[width= 12 cm]{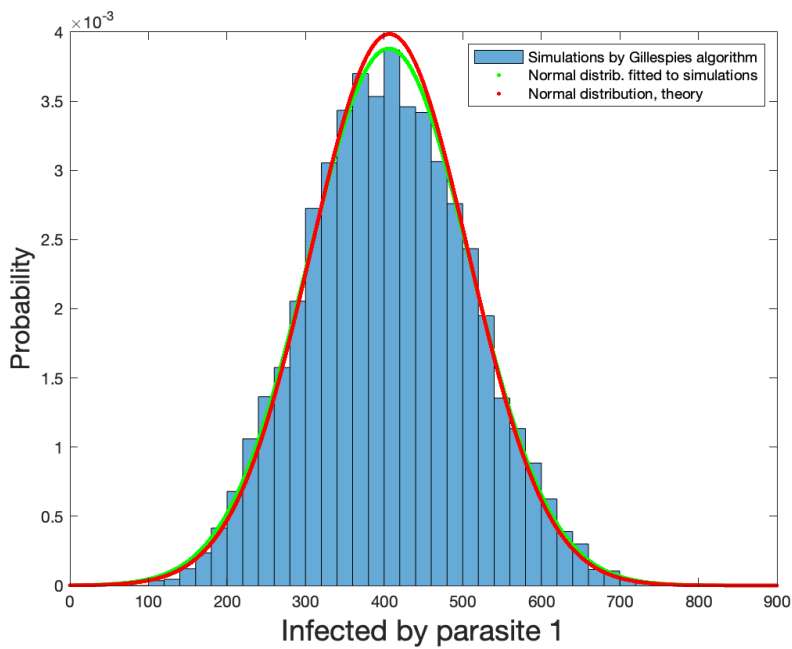}
\caption{Histogram generated from simulations using Gillespie's algorithm is compared to the probability density with mean and variance obtained from simulation using Gillespie's algorithm  and the probability density of normal distribution with mean and variance obtained from the theory of $I_1,$ infected by parasite 1 compartment at time = 300 of the stochastic model with quiescence. The initial population sizes of the model are; $I_1=50000, I_2=10000 ,Q_1=5000, Q_2= 5000$. The parameters of the model are $\beta_1 = \beta_2= 0.05, \Lambda = 1000,  d = 0.5,\nu_1= \nu_2 = 0.3, \gamma_{1} =  \gamma_{2} = 0.003,\zeta_1 = \zeta_2= 0.1, \rho_1 = \rho_2 =  0.7, \epsilon_1 = \epsilon_2 = 10.$}
\label{Fig2}
\end{figure}

In order to solve the above Fokker-Planck equation (\ref{FPeq}), we use the so-called asymptotic method (see for example \cite{ref3}). The principle is to transform the multivariate Fokker-Planck equation to a linear Fokker-Planck equation which is linearised around the stationary state of the deterministic system (\ref{Model1}). The solution of the linear Fokker-Planck is found to be normally distributed, the solution is given in the following two theorems (see chapter 8 of \cite{ref16}). We numerically checked this results using our stochastic simulations and the comparison is shown in (Figure \ref{Fig2}). \\

\newtheorem{mypro9}{Theorem}
\begin{mypro9}
The linear multivariate Fokker-Planck of (\ref{FPeq}) can be written as follows
\begin{equation}
\frac{\partial P(y,t)}{dt} = - \sum_{ij}^5 M_{ij} \frac{\partial}{\partial y_i} y_i P(y,t)+\frac{1}{2} \sum_{ij}^5 N_{ij} \frac{\partial^2}{\partial y_i \partial y_j}P(y,t) \\
\end{equation}
where $y = (y_1,\dots,y_5), N_{ij}$ is symmetric and positive definite, its solution is given as \\
$$P(y,t) = (2\pi)^\frac{1}{2} det (\Sigma)^\frac{1}{2} exp(-\frac{1}{2} y \Sigma^{-1} y^T) $$
with $$\Sigma^{-1}  = 2\int_0^\infty e^{-Mt} N e^{-Mt} dt. $$
\end{mypro9}
The matrices $N$ and $M$ are defined explicitly in Appendix D.
\begin{mypro9}
For every matrix $N$ which is symmetric and positive-definite, there a unique solution $\Sigma^{-1}$ to the following equation known as Lyapunov equation $$M \Sigma^{-1}+\Sigma^{-1} M^T  = N$$  where $\Sigma^{-1}$ is symmetric, positive-definite and equal to 
$$\Sigma^{-1}  = \int_0^\infty e^{-Mt} N e^{-M^Tt} dt. $$
\end{mypro9}
Theorem 2 which known as Lyapunov equation \cite{ref4} allows us to compute the covariance matrix as found in the normal distribution shown in theorem 1 fairly easily, this is due to the fact that matrices $A$ and $B$ are constant matrices, the only unknown is the $\Sigma^{-1}$ matrix. The covariance matrix is of dimension 5 and tells us the degree at which each compartments namely healthy, infected by strain 1 and 2 and quiescence class 1 and 2 go together \textit{i.e.} the relationship between each class. We use MATLAB to perform numerical calculations for the analytical solutions of the covariance matrix $\Sigma^{-1}$ .

We also computed 10,000 independent stochastic realisations using Gillespie's algorithm. The probability histogram was plotted in (Figure \ref{Fig2}) for the number of infected individuals by strain 1. This distribution is then compared with the probability density function of the normal distribution with mean and variance obtained from both Gilliespie's algorithm and the normal approximation method using linear multivariate Fokker-Planck equation (\ref{FPeq}). The results are consistent which further validates our analytical result obtained using linear Fokker-Planck.

\section{Covariance Matrix}
In order to understand the effect of quiescence in our stochastic model, we need to compare the system with quiescence to that of the system without quiescence in terms of the number of infected by both parasites. To do the comparative study we need to collapse the covariance matrix for both models with and without quiescence so that we only have $2 \time 2$ covariance matrix of the infected individuals. For the model with quiescence, this is done by adding the number of individuals in the infected class and the number of individuals in the quiescence stage to obtain a total number of infected individuals (irrespective of their quiescence status). For the system without quiescence, it is straight forward, it is achieved by isolating the number of individuals in the infected compartment. This step is justified below, and the following results indicate how to compute the covariance matrix (see \cite{ref5,ref23}). The obtained covariance matrix is denoted as the collapsed covariance matrix. \\
Let $\mathbf{Y} \sim \mathbf{N}_r (\mu, \Sigma)$ be r-variate multivariate normal distribution with mean $\mu$ and variance $\Sigma$, where 
 \begin{equation*} 
 \mathbf{Y} = \begin{bmatrix}
       Y_1         \\
       Y_2  \\
       \vdots  \\
       Y_{r}
     \end{bmatrix}
     \qquad
     \mu = \begin{bmatrix}
       \mu_1         \\
       \mu_2  \\
       \vdots  \\
       \mu_{r}
     \end{bmatrix}
     \qquad
     \Sigma = 
 \begin{bmatrix}
  \sigma_{1,1} & \sigma_{1,2} & \cdots & \sigma_{1,r} \\
  \sigma_{2,1} & \sigma_{2,2} & \cdots & \sigma_{2,r} \\
  \vdots  & \vdots  & \ddots & \vdots  \\
  \sigma_{r,1} & \sigma_{m,2} & \cdots & \sigma_{r,r} 
 \end{bmatrix}
 \end{equation*}
 Any q linear combination of the $Y_i$, say $\mathbf{A}'\mathbf{Y}, $ is (q-variate) multivariate normal. Let  \\
 \begin{equation*}
 \mathbf{A}' \mathbf{Y} = 
 \begin{bmatrix}
  a_{11}Y_1+a_{12} Y_2 + \dots +a_{1r} Y_r\\ \\
  a_{21} Y_1 + a_{22} Y_2 + \dots + a_{2r}  Y_r\\ \\
  \dots  +  \dots  +  \dots  +  \dots \\ \\
  a_{q1} Y_1+a_{q2} Y_2 + \dots + a_{qr} Y_r
 \end{bmatrix},
 \end{equation*}
 then 
 \begin{equation}
  \mathbf{A}' \mathbf{Y} \sim \mathit{N}_q(\mathbf{A}' \mu, \mathbf{A}' \Sigma \mathbf{A}).
  \end{equation}

Numerical examples of the collapsed covariance matrix are shown for various parameter combinations. The collapsed covariance matrix of the model with quiescence is denoted as $\mathnormal{E_q}$ and the collapsed covariance matrix of the model without quiescence as $\mathnormal{E_wq}$. In an effort to understand the effect of quiescence on the stochastic process, we consider two different cases of parameter combinations: symmetric where the parameter values of stain 1 and 2 are exactly the same (examples 1, 2, and 3), and non-symmetric where the parameter values of stain 1 and 2 are different (for example $\rho_1 \ne \rho_2$, examples 4, 5, 6 and 7). 
\bigskip

\textbf{Example 1}
We fix the following parameter values: $\beta_1 = \beta_2 = 0.005, d = 0.5, \Lambda = 1000, \nu_1 = \nu_2 = 0.3, \rho_1 = \rho_2 = 0.7, \gamma_1=\gamma_2 = 0.003, \zeta_1 = \zeta_2 =0.1, \epsilon_{1} = \epsilon_2 = 0.6$ and the initial population sizes are $H =50,000,$ $I_1=10,000, I_2=10,000, Q_1 = 5,000, Q_2 = 5,000, \text{time} = 300.$ We obtain the following collapsed covariance matrices:
$$
\mathnormal{E_{q1}} = 
\begin{pmatrix}
683,640    & -682,500 \\
-682,500   &   683,640 
\end{pmatrix},
\qquad
\mathnormal{E_{wq1}} = 
\begin{pmatrix}
298,630 & -297,560 \\
-297,560 &  298,630
\end{pmatrix}.$$
\medskip  

\textbf{Example 2}
We use the same parameter values as in example 1 only with a lower quiescence rate $\rho_1 = \rho_2 = 0.4 $
$$
\mathnormal{E_{q2}} = 
\begin{pmatrix}
655,170    & -654,060 \\
-654,060   &   655,170 
\end{pmatrix},
\qquad
\mathnormal{E_{wq2}} = E_{wq1} $$ 
We show in example 1 that the model with quiescence exhibits a larger variance compared with the model without quiescence. When comparing example 1 and 2, we observe the effect of quiescence on reducing the variance of the number of infected individuals. When the rate of entering quiescence stage ($\rho$) decreases, the variance of the number of infected individuals decreases ($E_{q1}$ versus $E_{q2}$).
\bigskip

\textbf{Example 3}
The parameter and initial values are identical to example 1 except that the disease transmission rates are now 10 times lower $\beta_1 = \beta_2 = 0.0005$: 
$$
\mathnormal{E_{q3}} = 
\begin{pmatrix}
14.81    & -0.0388 \\
-0.0388   &   14.81
\end{pmatrix},
\qquad
\mathnormal{E_{wq3}} = 
\begin{pmatrix}
27,651 & -26,443 \\
-26,443 &  27,651
\end{pmatrix}.$$
In example 3, we observe the effect of decreasing the transmission rate in reducing the variance and covariance of the collapsed covariance matrix. In contrast to example 1, in example 3, we find that the model with quiescence has less variance compared to the model without quiescence. 
\bigskip

\textbf{Example 4}
We use the same parameter values as in example 1 only with asymmetric rates of quiescence $\rho_1 = 0.3,\rho_2 = 0.5 $
$$
\mathnormal{E_{q4}} = 
\begin{pmatrix}
2,251.9    & -57.42 \\
-57.42   &   64.35 
\end{pmatrix},
\qquad
\mathnormal{E_{wq4}} = E_{wq1} $$
Now that we use asymmetrical rates of entering quiescence between the two strains in example 4, the variance are much decreased compared to examples 1 and 2. This further reduction in variance occurs because of the competition amongst the two parasite types in the model with quiescence (which was absent because of symmetrical rates in examples 1-3). In other words, because the two parasite strains have different quiescence rates, there is also competition between them to infect host individuals. Furthermore, the strain with the largest rate of entering the quiescence stage ($\rho$) exhibits a smaller variance than the strain with a lower quiescent rate. By analogy, we call this phenomenon as moving average behaviour (see discussion). 
\bigskip

\textbf{Example 5}
We use the same parameter values as in example 1 only with asymmetric rates of entering $\rho_1 = 0.8, \rho_2 = 0.4 $ and exiting $\zeta_1 = 0.4, \zeta_2 =0.8$ quiescence. 
$$
\mathnormal{E_{q5}} = 
\begin{pmatrix}
19.17    & -15.07 \\
-15.07   &   2187.1 
\end{pmatrix},
\qquad
\mathnormal{E_{wq5}} =  E_{wq1}.$$
In example 5, we investigate the influence of asymmetric rates of entering and exiting the quiescent stage on the variance in infected individuals. We set the rate of entering quiescence of strain 1 to be larger than rate of strain 2, while the rate of exiting quiescence of strain 1 is smaller than that of strain 2. We still observe the so-called moving average effect, that is, the strain with the largest rate of entering the quiescence has the smaller variance. This example shows that entering quiescence has significant effect in changing the dynamics of the system. 
\bigskip

\textbf{Example 6}
We use the same parameter values as in example 1 only with asymmetric rates of entering $\rho_1 = 0.8, \rho_2 = 0.4 $ and exiting $\zeta_1 = 0.8, \zeta_2 =0.4$ quiescence. 
$$
\mathnormal{E_{q6}} = 
\begin{pmatrix}
164.04    & -151.92 \\
-151.92   &   2332.6 
\end{pmatrix},
\qquad
\mathnormal{E_{wq6}} =  E_{wq1}.$$
In example 6, we take the rate of entering and exiting quiescence to be the same for each strain,that is, $\rho_1 = 0.8 = \zeta_1 = 0.8, \rho_2 = 0.4 = \zeta_2 =0.4, $ to ascertain if the moving average is determined by the rate of entering quiescence or the longest quiescence time. This example confirms that the moving average is determined by the rate of entering quiescence. We note by this example that rate of exiting quiescence stage doesn't effect the dynamic significantly as far as the moving average is concern. 
\bigskip

\textbf{Example 7}
In example 7, we increase the disease transmission rates and decrease the birth and death rate (compared to example 1), while we assume asymmetric rates of entering quiescence (as in example 5) but symmetric rates of exiting quiescence as well as the immigration rate. The following values are used $\beta_1 = \beta_2 = 0.05, d = 0.4, \Lambda = 100, \nu_1 = 0.03, \nu_2 = 0.3, \rho_1 = 0.8, \rho_2 = 0.4, \gamma_1=\gamma_2 = 0.03, \zeta_1 = \zeta_2 =0.1, \epsilon_1 =\epsilon_2 = 0.6$ and the initial population sizes are as in example 1. We obtain the following collapsed covariance matrices: 
$$
\mathnormal{E_{q6}} = 
\begin{pmatrix}
967.63 & -927.22 \\
-927.22 & 1151.1
\end{pmatrix},
\qquad
\mathnormal{E_{wq6}} = 
\begin{pmatrix}
245.56 & -3.6384 \\
-3.6384  & 5.8915
\end{pmatrix}.$$ 
From examples 7, here we use asymmetric values of parameters in both models, we see the influence of quiescence in reducing the variance of the collapsed covariance matrix whenever one of the rates of entering quiescence is high. In addition, we also see the effect of strain competition in the model without quiescence in reducing the variance of the number of infected individuals. In the model with quiescence we take the recovery rate of infected individuals by strain 1 to be 10 times smaller than those infected by strain 2, and observe our moving average effect.

\begin{figure}[h!]
    \begin{subfigure}[b]{0.55\textwidth}
        \includegraphics[width=\textwidth]{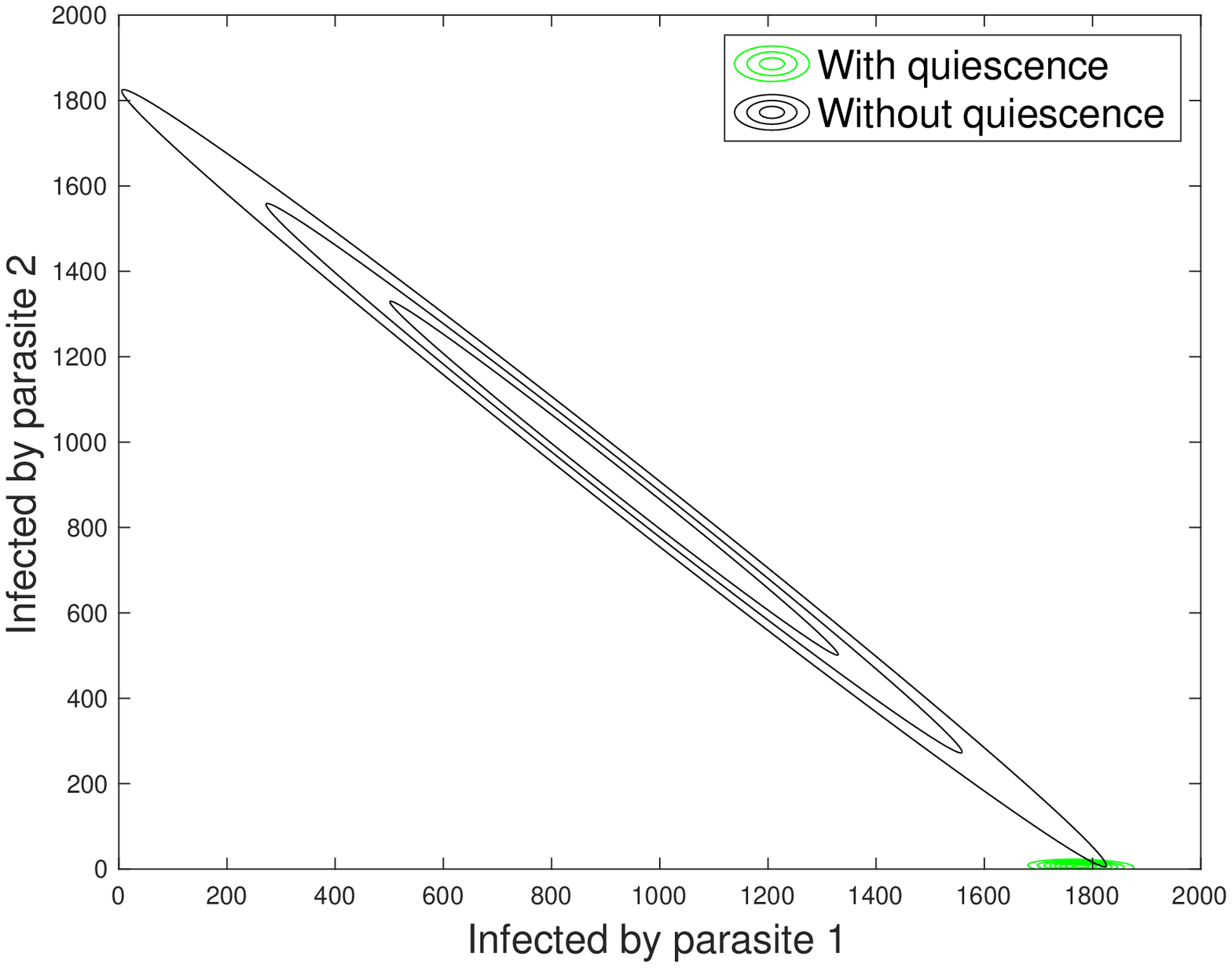}
        \caption{}
        \label{conex4}
    \end{subfigure}
    ~ 
    \begin{subfigure}[b]{0.55\textwidth}
        \includegraphics[width=\textwidth]{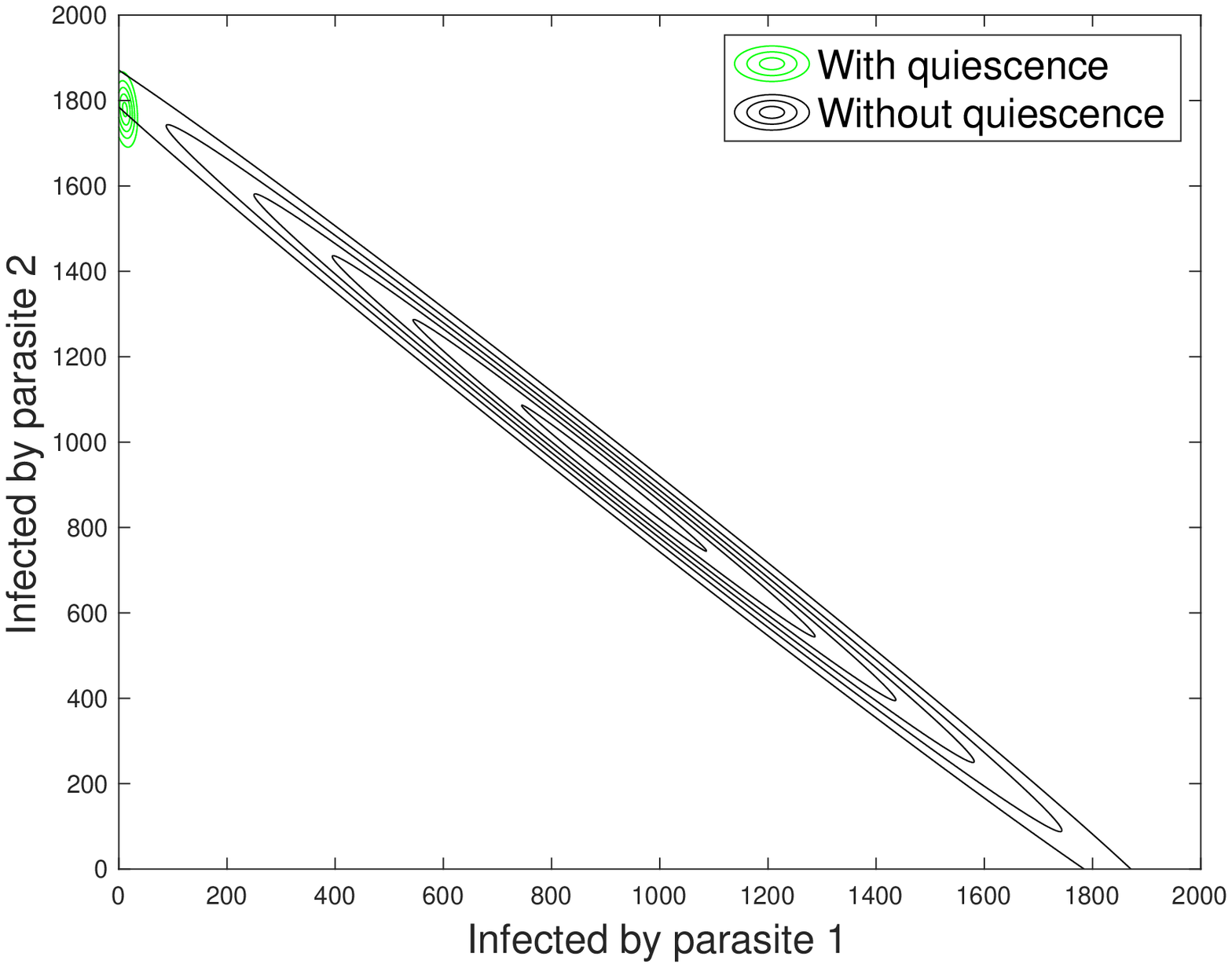}
        \caption{}
        \label{conex6}
    \end{subfigure}
    \caption{Contour plots of the joint density of infected individuals by strain 1 and 2 based on simulations for (a) example 4, and (b) example 5 considered in the text. The x-axis is the number of infected individuals of strain 1 while the y-axis is the number of infected individuals by strain 2 based on the parameters stated in each example.}
\end{figure}

\begin{figure}[h!]
    \begin{subfigure}[b]{0.55\textwidth}
        \includegraphics[width=\textwidth]{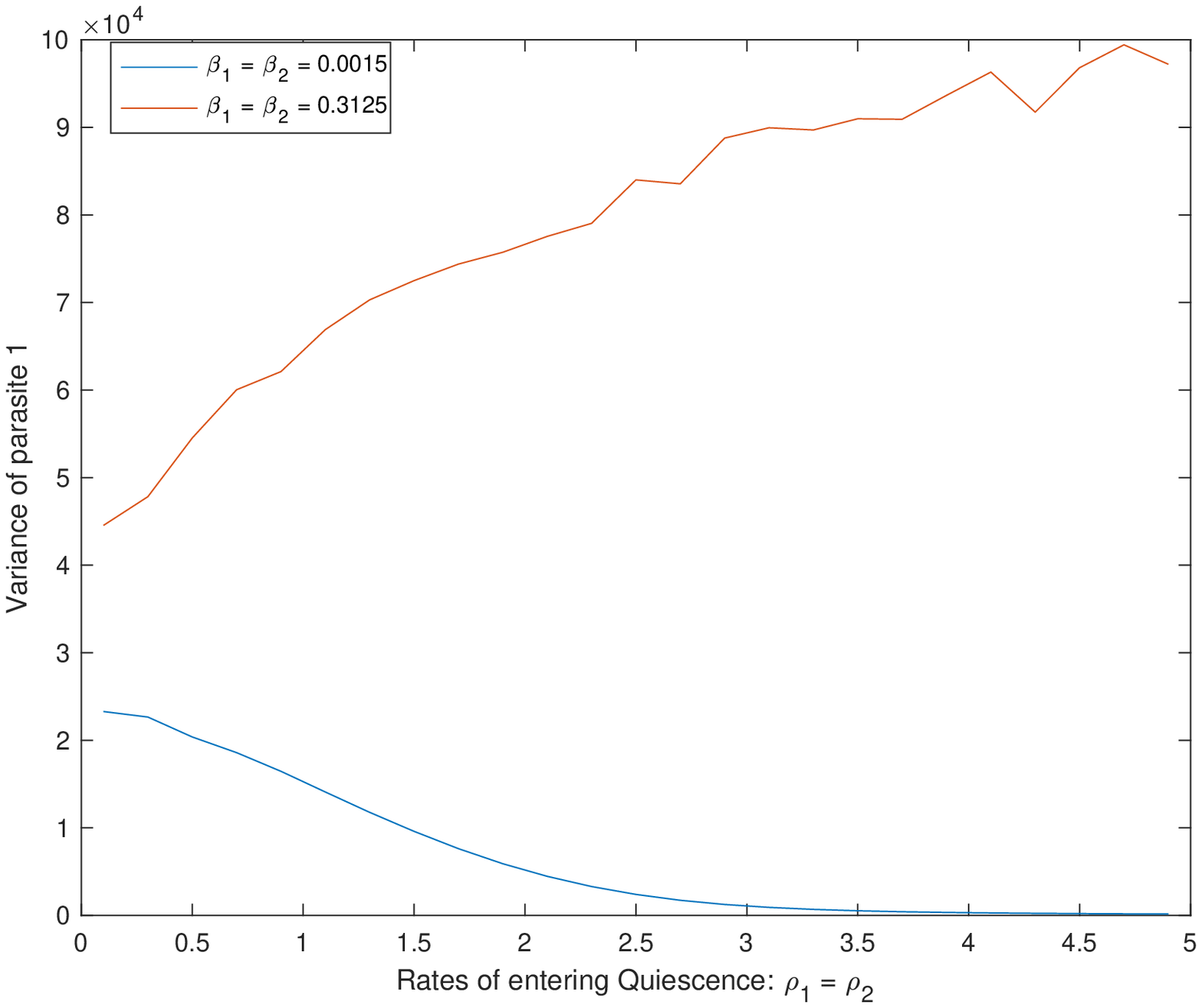}
        \caption{}
        \label{var}
    \end{subfigure}
    ~ 
    \begin{subfigure}[b]{0.55\textwidth}
        \includegraphics[width=\textwidth]{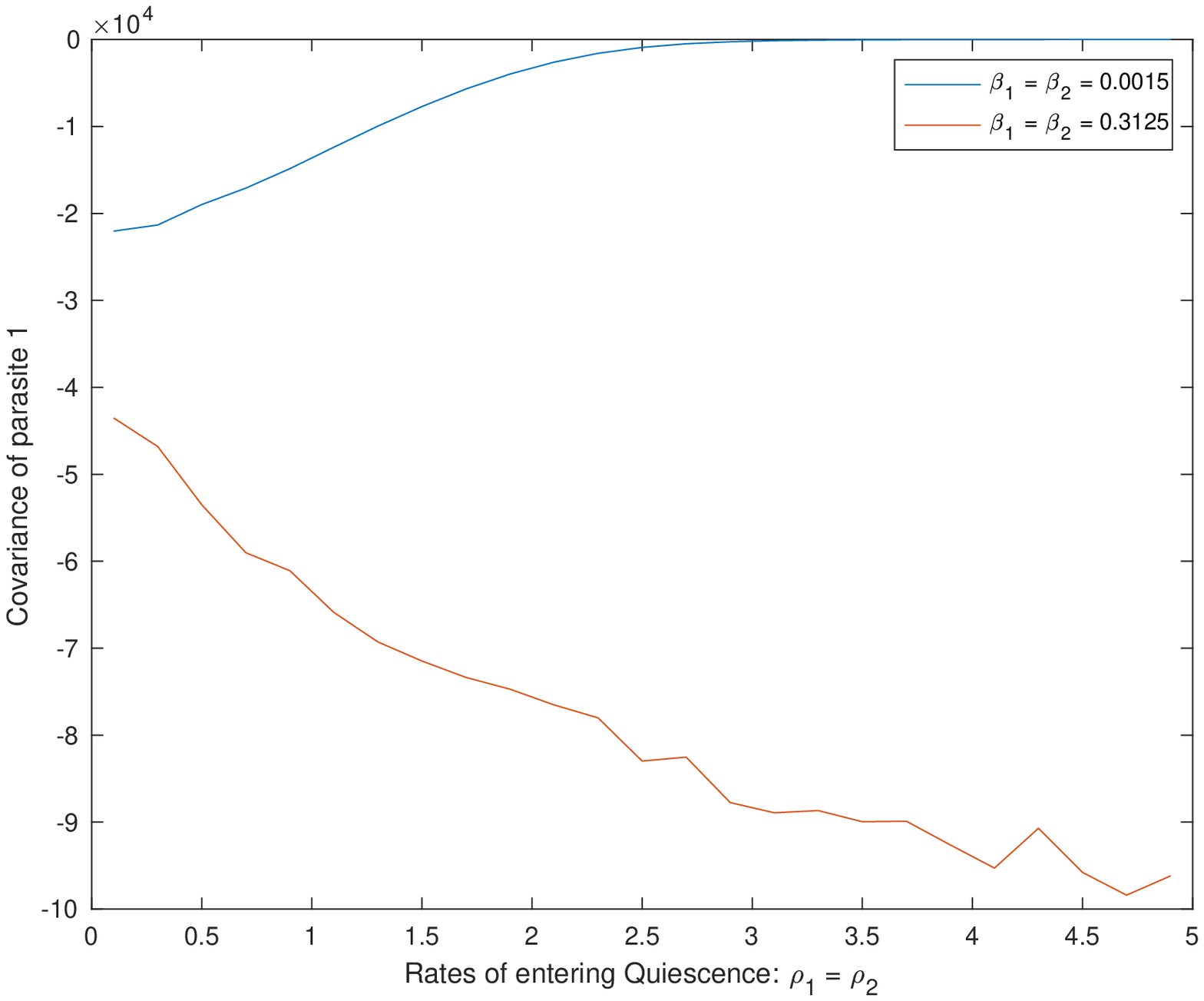}
        \caption{}
        \label{cov}
    \end{subfigure}
    \caption{Effect of quiescence, rates of entering the quiescence phase $\rho_1 = \rho_2$, and of transmission rates $\beta_1 = \beta_2$ on the (a) variance of parasite 1, and (b) covariance of parasite 1 of the collapsed covariance matrix. We use the following parameter values (symmetrical case): $ d = 0.5, \Lambda = 1000, \nu_1 = \nu_2 = 0.3, \gamma_1=\gamma_2 = 0.003, \zeta_1 = \zeta_2 =0.1, \epsilon_{1} = \epsilon_2 = 10 $ and the initial population sizes are $H =50,000,$ $I_1=10,000, I_2=10,000, Q_1 = 5,000, Q_2 = 5,000, \text{time} = 300.$ The blue line is for $\beta_1= \beta_2= 0.0015$, and the red line for $\beta_1=\beta_2=0.3125$}.

\end{figure}

As additional verification, we draw contour plots of the joint density of infected individuals by strain 1 and 2 in (Figure \ref{conex4}) and (Figure \ref{conex6}) which compare the variance in the number of infected individuals by both strains. We confirm that the joint distribution of the number of infected individuals by parasite strain 1 and 2 have a smaller surface area, that is with less variance, under the model with quiescence than the absence of quiescence. In all examples, the values of the covariance (off-diagonal elements) are negative, and we observe this effect also in the contours (Figures \ref{conex4}, \ref{conex6}) because the number of infected individuals by parasite 1 and 2 are negatively correlated. This negative correlation is a result of the competition between the parasite types. We finally analyse the change in variance (Figure \ref{var}) and covariance (Figure \ref{cov}) of the collapsed covariance matrix as a function of $\rho_1$  and $\rho_2$ (rates of entering quiescence). The effect of the transmission rates $\beta_1$ and $\beta_2$ is here again visible: when $\beta_1 = \beta_2$ are low, high rates of entering quiescence depletes the infected compartments so that the number of infected drops down and the infection decreases, which in turn reduces the variance. When $\beta_1 = \beta_2$ are high, there are enough infected to keep the infection spreading despite the rate of quiescence, hence the increases in the variance (under a fixed values of $\zeta_1$ and $\zeta_2$ (Figures \ref{var},\ref{cov}). The behaviour of the covariance is reversed as the infected classes are negatively correlated. Based on the examples above, increasing $\zeta_1$  and $\zeta_2$ would results in decreasing the difference between the variance (as well as for the covariance) for the different transmission rates $\beta_1$ and $\beta_2$.

\section{Discussion}

In this study we aim to understand the effect of quiescence on the spread of infectious disease and with competition between parasite strains. Our study shows that introducing the pathogens ability to switch between an active and inactive (quiescence) phase can significantly impact the stochasticity in the system. In our system, when the invasion/immigration rates are turned off, one of the parasite type becomes extinct. However, when the invasion/immigration rates are turned on, coexistence of host and both parasite types is possible. If both strains show equal rates of infection, transmission and quiescence, there is no real competition and the system behaves as if only one parasite would be present. On other hand, when the parasite types have different characteristics, there is competition between them which generates various epidemiological dynamics.

Our collapsed covariance measure quantifies the infection load at the steady state of the system with and without quiescence. We measure this infection load for various parameter combinations of interest to understand the impact of quiescence on the stochastic process. Under symmetric quiescence rates and high transmission rates, quiescence increases the variance in infected individuals, while the quiescence reduces the variance in infected when transmission rates are low. When considering asymmetry in quiescence rates between parasite strains, we uncover a special phenomenon which we call by analogy to the moving average behaviour. The strain with the high rate of entering quiescence serves as moving average for the whole parasite population and buffers the effect of the second less quiescent strain. In other words, quiescence reduces the intensity of the noise in the stochastic infection process determining the variance of the number of infected individuals. Moving average is a well known concept in sound, signal, and image processing. In sound processing for example, moving average also known as low pass filter, filters the frequencies so that only low frequencies can be heard. The sound of noisy wave or distorted signal, is being smoothens by applying a moving average processing function because it assumes the areas of high frequencies as noise. We are not aware of the use of moving average in the field of disease epidemiology, and hence introduce it here as a consequence of quiescence in parasite. When different strains of parasite do show different quiescent rates, the competition between them under a stochastic epidemiological process reduces the number of infected individuals, as well as the virulence of the disease (number of host death). We theoretically predict that under competition between parasite types, the strain with the lower rate of entering quiescence gets fixed, however, if coexistence can be maintained by influx of parasite strains from outside, quiescence has the beneficial effect to reduce the stochasticity of the system. An extension for our work is to investigate if quiescence itself can evolve in such epidemiological setup as a bet-hedging strategy reducing stochasticity in transmission rates.

Due to the difficulty in the existing methods to analyse the stability of $5 \times 5$ matrix, we developed here a criterion for the study of stability of the system with quiescence for the deterministic system. Proposition 2 is important because it reduces the dimension of the system from 5 to 3. It is well known that studying the stability of the system with higher dimension is hard, often times impossible. While system with low dimension is easy and straight forward to study its stability. Thus the reduction in proposition 2 is of significant importance that removes the difficulties of analysing matrix with high dimension.

We then extended our model to a stochastic version. We show that the analytic solution of the linear Fokker-Planck equation is normally distributed with mean around the equilibrium solution. We confirm this results by computing 10,000 independent stochastic realisations using Gillespie's algorithm (Figure \ref{Fig2}). The probability histogram was plotted at a time equals to 300 generations. This distribution is then compared with the probability density function of the normal distribution with mean and variance as obtained from both Gilliespie's algorithm and the normal approximation method using linear multivariate Fokker-Planck equation (\ref{FPeq}). The results are consistent which further validates our analytical result obtained using the linear Fokker-Planck equation.

As revealed by a wealth of recent studies on plant or animal, microbiomes are composed of multiple species and multiple strains per species. The composition of species and/or strains is governed by antagonistic, mutualistic or neutral inter- and intra-specific interactions along with stochastic processes such as birth and death, extinction-recolonization and migration of strains/species [see \cite{ref12,ref19}]. We speculate that our results on quiescence should be affecting the dynamics in these multi-species systems. Moreover, many microbe, especially human parasites, enter quiescence stage as a mechanism of resistance against antibiotics \cite{ref48}. This has important consequences for the management of infectious diseases. Furthermore, host bacteria can also enter quiescence upon contact with viruses \cite{ref50}, which can lead to changes in the expected population dynamics of the bacterial and virus populations \cite{ref52}. It is therefore of paramount importance to understand the influence of the quiescence on the population of hosts and parasites, especially as coevolution between antagonistic species can drive the evolution of quiescence/dormancy \cite{ref42}. We show here that if quiescence reduces stochasticity and reduces the noise under strain competition, the same idea should be investigated for a model of bacteria submitted to stochasticity of antibiotic treatment. We speculate that quiescence is not only a bet-hedging strategy, but also influences the stochasticity of the population behaviour, namely the population size of bacteria becoming more stable in time and insensitive to antibiotic treatment.
\newpage
\bibliographystyle{plain} 
\bibliography{qref}

\newpage
\textbf{Appendix A, Equilibrium Solution of the Model with Quiescent} \\
From equations 5 and 4 of system (\ref{Model1}), the quiescence compartments, we find the equilibrium solutions and is given as follows \\
$Q_1^* = \frac{\rho_1 I_1}{\zeta_1+d,}$  and $Q_2^* = \frac{\rho_2 I_2}{\zeta_2+d},$
let $c_1 = \frac{\rho_1}{\zeta_1+d}, c_2 = \frac{\rho_2}{\zeta_2+d},$ then the equilibrium solutions of the infected compartment (equations 1 and 2 of system \ref{Model1}) are given by
$$I_1^*= \frac{\epsilon_1}{d+\gamma_1+\nu_1+\rho_1-\zeta_1c_{11}-\beta_1H^*}, I_2^*= \frac{\epsilon_2}{d+\gamma_2+\nu_2+\rho_2-\zeta_2c_{12}-\beta_2H^*}.$$\\
Now we need to calculate the equilibrium solution in the healthy compartment, to do so we need the following propositions.
\newtheorem{my1}{Proposition}
 \begin{my1} 
For $\epsilon_1, \epsilon_2 > 0,$ there is at least one non-negative equilibrium solution in the healthy compartment.
  \end{my1}
 \begin{proof}
  Substituting the equilibrium solutions of the quiescence and infected compartments as calculated above in the first equation of the system (\ref{Model1}), we have
  
 $P(H) = \Lambda(d+\gamma_1+\nu_1+\rho_1-\zeta_1c_{1}-\beta_1H)(d+\gamma_2+\nu_2+\rho_2-\zeta_2c_{2}-\beta_2H)-\beta_1H\epsilon_1(d+\gamma_2+\nu_2+\rho_2-\zeta_2c_{2}-\beta_2H)-\beta_2H\epsilon_2(d+\gamma_1+\nu_1+\rho_1-\zeta_1c_{1}-\beta_1H)-dH(d+\gamma_1+\nu_1+\rho_1-\zeta_1c_{1}-\beta_1H)(d+\gamma_2+\nu_2+\rho_2-\zeta_2c_{2}-\beta_2H)+\nu_1\epsilon_1(d+\gamma_2+\nu_2+\rho_2-\zeta_2c_{2}-\beta_2H)+\nu_2\epsilon_2(d+\gamma_1+\nu_1+\rho_1-\zeta_1c_{1}-\beta_1H),$
  then
  $P(0) = \Lambda(d+\gamma_1+\nu_1+\rho_1-\zeta_1c_{1})(d+\gamma_2+\nu_2+\rho_1-\zeta_1c_{1})+\nu_1\epsilon_1(d+\gamma_2+\nu_2+\rho_2-\zeta_2c_{2})+\nu_2\epsilon_2(d+\gamma_1+\nu_1+\rho_1-\zeta_1c_{1})> 0,$ because the terms inside brackets are all positive. \\
  and \\
$P(H) \rightarrow -\infty, $ then by intermediary value theorem there exist $H^* $ such that $$P(H^*) = 0, H^*>0$$
  Please observe that other compartments $(I_1^*,I_2^*,Q_1^*,Q_2^*)$ for $H^*$ are non-negative, since
 $$P\Big(\frac{d+\gamma_1+\nu_1+\rho_1 -\zeta_1c_{1}}{\beta_1} \Big) < 0, \implies H^* \leq \frac{d+\gamma_1+\nu_1+\rho_1 -\zeta_1c_{1}}{\beta_1} \implies I_1^* \geq 0,$$
  by the same argument, we show that $I_2^* >0.$ Since $I_1^*, I_2^*>0,$ then $Q_1^*,Q_2^*> 0$
 \end{proof} 
In the above proposition 1, we find a polynomial of degree three in which we use intermediate value theorem to show that the polynomial has a solution. \\ 
 
\textbf{Uniqueness of The Equilibrium Solution} \\
We introduce the terms $a,b,c,d$ defined bellow, with this notation, we obtain the following proposition
 \newtheorem{myt3*}{Proposition}
 \begin{myt3*}
 If $b^2 < 3ac$ , then there is a unique non-negative equilibrium solution of $P(H).$ 
 \end{myt3*}
 \begin{proof}
 Let 
 $$P(H) = aH^3+bH^2+cH+d = 0,$$
 \begin{equation}
 \frac{dP}{dH} = 3aH^2+2bH^2+c = 0.
 \label{poly}
 \end{equation}
 The solution of quadratic equation (\ref{poly}) is 
 \begin{equation}
 H = \frac{-(2b) \pm \sqrt{(2b)^2-4(3a)c}}{2(3a)}
 \label{qua}
\end{equation}
 
  where $$a = -3\beta_1\beta_2d, $$
 $b = 2d\beta_1\rho_2+2d\beta_2\rho_1+2d\beta_1\nu_2+2d\beta_1\nu_1-2 c_{12}d\beta_1\zeta_2-2 c_{11}d\beta_2\zeta_1+2\beta_1\beta_2\epsilon_2+2\beta_1\beta_2\epsilon_1+ 2d\beta_1\gamma_2+2d\beta_1\gamma_1+2 \Lambda\beta_1\beta_2+2 d^2  \beta_2+2 d^2 \beta_1,$ \\ 
 $c = -\beta_1\epsilon_1\nu_2-\Lambda \beta_1\nu_2-\beta_2\epsilon_1\nu_1-\Lambda\beta_2\nu_1-d\rho_1\rho_2- d\nu_1\rho_2+c_{11}d\zeta_1\rho_2-\beta_1\epsilon_1\rho_2-d\gamma_1\rho_2-\Lambda\beta_1\rho_2-d^2\rho_2-d\nu_2\rho_1+c_{12}d\zeta_2\rho_1-\beta_2\epsilon_2\rho_1-d\gamma_2\rho_1-\Lambda \beta_2\rho_1-d^2\rho_1-d\nu_1\nu_2+c_{11}d\zeta_1\nu_2-\beta_1\epsilon_2\nu_2-d^2\nu_2+c_{12}d\zeta_2\nu_1-\beta_2\epsilon_2\nu_1-d\gamma_2\nu_1-d^2\nu_1-c_{11}c_{12}d\zeta_1\zeta_2+c_{12}\beta_1\epsilon_1\zeta_2+c_{12}d\gamma_1\zeta_2+c_{12}\Lambda\beta_1\zeta_2+c_{12}d^2\zeta_2+c_{11}\beta_2\epsilon_2\zeta_1+c_{11}d\gamma_2\zeta_1+c_{11}\Lambda\beta_2\zeta_1+c_{11}d^2\zeta_1-\beta_2\gamma_1\epsilon_2-d\beta_2\epsilon_2-\beta_1\gamma_2\epsilon_1-d\beta_1\epsilon_1-d\gamma_1\gamma_2-\Lambda\beta_1\gamma_2-d^2\gamma_2-\Lambda\beta_2\gamma_1-d^2\gamma_1-\Lambda d \beta_1-d^3,$ \\
choose parameter values so that 
 $$b^2 < 3ac ,$$
then the quadratic equation (\ref{qua}) does not have real solution. 
\end{proof}
In the above proof, we use calculus to find the maximum value of the polynomial. The analysis shows that the polynomial does not have a maximum or minimum value at the specified interval. This shows that the polynomial has only one root by proposition 1 (existence of a solution) above.
\newpage
\textbf{Appendix B: Proof of proposition 2 stated in section 2.3} \\
We now proof proposition 2 stated in section 2.3 above regarding the stability of the matrix $B$ defined in \ref{matrixB}. 
\begin{proof}
The characteristics polynomial of $B$ is given by 
$$ \lambda^5+b_1\lambda^4+b_2\lambda^3+b_3 \lambda^2+b_4 \lambda+b_5 = 0 $$
where 
 \begin{equation*}
 \begin{aligned}
 & b_1 = \rho_1+\rho_2+\zeta_1+\zeta_2-\text{tr(A)}\\ 
 & b_2 = \rho_1 \rho_2+\rho_1 \zeta_1+\rho_2 \zeta_1+\zeta_1 \zeta_2-\zeta_1 \text{tr(A)}-\zeta_2 \text{tr(A)}-(a_{11}+a_{33}) \rho_2-(a_{22}+a_{33}) \rho_1+a_2\\ 
 & b_3 =\zeta_1a_2+\zeta_2a_2+(a_{11} a_{33}-a_{13} a_{31}) \rho_2+(a_{22} a_{33}-a_{23} a_{32}) \rho_1 -\text{det(A)}-\zeta_1\zeta_2 \text{tr(A)} -(a_{22}+a_{33}) \rho_1 \zeta_2 \\
 & -a_{33} \rho_1 \rho_2-a_{33} \rho_2 \zeta_1-a_{11} \rho_2 \zeta_1\\ 
 & b_4 =\zeta_1 \zeta_2 a_2+(a_{22} a_{33}-a_{23} a_{32}) \rho_1 \zeta_2 +(a_{11} a_{33}-a_{13} a_{31}) \rho_2 \zeta_1  -(\zeta_1 +\zeta_2 )\text{det(A)}\\ 
 & b_5 = - \zeta_1 \zeta_2 \text{det(A)}
 \end{aligned}
 \end{equation*}
 \textit{Step 1:} \\ 
By Routh-Hurwitz Criterion \cite{ref1,ref39,ref40} , the matrix $B$ is stable if and only if the following conditions hold:   
\begin{enumerate}[label=(\roman*)]
\item $b_i > 0, \quad (i = 1, \dots, 5)$ \\ 
\item $b_1 b_2 b_3 > b_3^2+b_1^2 b_4$ \\
\item $(b_1 b_4-b_5)(b_1 b_2 b_3-b_3^2-b_1^2 b_4) > b_5 (b_1 b_2-b_3)^2+b_1 b_5^2$ \\ 
\textit{Step 2} \\ 
Suppose that for all $\rho_1, \rho_2, \zeta_1, \zeta_2 > 0 $ 
 \item $  b_1 > 0, \\ 
  = \rho_1+\rho_2+\zeta_1+\zeta_2-\text{tr(A)} > 0  \implies \text{tr(A)} \le 0$ \\ 
 \item $b_2   > 0  \\ 
 = \rho_1 \rho_2+\rho_1 \zeta_1+\rho_2 \zeta_1+\zeta_1 \zeta_2-\zeta_1 \text{tr(A)}-\zeta_2 \text{tr(A)}-(a_{11}+a_{33}) \rho_2-(a_{22}+a_{33}) \rho_1+a_2 > 0 \\ 
 \implies \text{tr(A)} \le 0, \quad a_{11} \le 0, \quad a_{22} \le 0, \quad \text{and} \quad a_{33} \le 0 $\\  
 \item $b_3 > 0 \\
 =\zeta_1a_2+\zeta_2a_2+(a_{11} a_{33}-a_{13} a_{31}) \rho_2+(a_{22} a_{33}-a_{23} a_{32}) \rho_1 -\text{det(A)}-\zeta_1\zeta_2 \text{tr(A)} -(a_{22}+a_{33}) \rho_1 \zeta_2 > 0\\ 
 \implies \text{det(A)} < 0,\quad  \text{tr(A)} \le 0, \quad a_{11} \le 0, \quad a_{22} \le 0, \\ 
  \quad a_{33}\le 0, \quad a_{13} a_{31} \le a_{11} a_{33} ,\quad \text{and} \quad  a_{23} a_{32} \le a_{22} a_{33} $ \\ 
 \item   $b_4 >0    \\
 \implies \zeta_1 \zeta_2 a_2+(a_{22} a_{33}-a_{23} a_{32}) \rho_1 \zeta_2 +(a_{11} a_{33}-a_{13} a_{31}) \rho_2 \zeta_1 > (\zeta_1 +\zeta_2 )\text{det(A)} $\\ 
$ \implies \text{det(A)} < 0, \quad a_{13} a_{31} \le a_{11} a_{33} ,\quad \text{and} \quad  a_{23} a_{32} \le a_{22} a_{33} $ \\
\item $b_5>0  \\ 
  = - \zeta_1 \zeta_2 \text{det(A)} > 0 \implies \text{det(A)} < 0$ \\
\textit{Step 3:} \\ 
Assume that 

$ \text{det(A)} < 0,\quad  \text{tr(A)} \le 0,\quad a_2 > 0, \quad a_{11} \le 0, \quad a_{22} \le 0, \\
  a_{33}\le 0, \quad a_{13} a_{31} \le a_{11} a_{33} , \quad  a_{23} a_{32} \le a_{22} a_{33} . $
  Then for all $\rho_1, \rho_2, \zeta_1, \zeta_2 > 0,$ we have \\
   \item $\rho_1+\rho_2+\zeta_1+\zeta_2-\text{tr(A)}  = b_1 > 0$  \\
   \item  $\rho_1 \rho_2+\rho_1 \zeta_1+\rho_2 \zeta_1+\zeta_1 \zeta_2-\zeta_1 \text{tr(A)}-\zeta_2 \text{tr(A)}-(a_{11}+a_{33}) \rho_2-(a_{22}+a_{33}) \rho_1+a_2  = b_2 >  0 $\\ 
   \item $  \zeta_1a_2+\zeta_2a_2+(a_{11} a_{33}-a_{13} a_{31}) \rho_2+(a_{22} a_{33}-a_{23} a_{32}) \rho_1 -\text{det(A)}-\zeta_1\zeta_2 \text{tr(A)} -(a_{22}+a_{33}) \rho_1 \zeta_2 -a_{33} \rho_1 \rho_2-a_{33} \rho_2 \zeta_1-a_{11} \rho_2 \zeta_1   = b_3 > 0 $ \\ 
   \item $ \zeta_1 \zeta_2 a_2+(a_{22} a_{33}-a_{23} a_{32}) \rho_1 \zeta_2 +(a_{11} a_{33}-a_{13} a_{31}) \rho_2 \zeta_1  -(\zeta_1 +\zeta_2 )\text{det(A)}  = b_4 > 0 $\\ 
   \item $-\zeta_1 \zeta_2 \text{det(A)}  = b_5 > 0$ \\
  \item 
  \begin{equation}
 \begin{aligned}
 & (\rho_1+\rho_2+\zeta_1+\zeta_2-\text{tr(A)})( \rho_1 \rho_2+\rho_1 \zeta_1 +\rho_2 \zeta_1+\zeta_1 \zeta_2-\zeta_1 \text{tr(A)}-\zeta_2 \text{tr(A)} \\
 &-(a_{11}+ a_{33}) \rho_2-(a_{22}+a_{33}) \rho_1+a_2 ) (-\text{det(A)}+\zeta_1a_2+\zeta_2a_2+(a_{11} a_{33}-a_{13} a_{31}) \rho_2 \\
 & +(a_{22} a_{33}-a_{23} a_{32}) \rho_1-(a_{22}+a_{33}) \rho_1 \zeta_2  -a_{33} \rho_1 \rho_2-a_{33} \rho_2 \zeta_1-a_{11} \rho_2 \zeta_1-\zeta_1\zeta_2 \text{tr(A)} ) \\
& -(-\text{det(A)}+\zeta_1a_2+\zeta_2a_2+(a_{11} a_{33}-a_{13} a_{31}) \rho_2 +(a_{22} a_{33}-a_{23} a_{32}) \rho_1-(a_{22}+a_{33}) \rho_1 \zeta_2\\ 
 & -a_{33} \rho_1 \rho_2-a_{33} \rho_2 \zeta_1-a_{11} \rho_2 \zeta_1-\zeta_1\zeta_2 \text{tr(A)} )^2 -(\rho_1+\rho_2+\zeta_1+\zeta_2-\text{tr(A)})^2(-\zeta_1 \text{det(A)} \\
 & -\zeta_2\text{det(A)} +\zeta_1 \zeta_2 a_2 +(a_{22} a_{33}-a_{23} a_{32}) \rho_1 \zeta_2 +(a_{11} a_{33}-a_{13} a_{31}) \rho_2 \zeta_1 ) 
 \end{aligned}
 \label{ctritera1}
 \end{equation}
  = $b_1 b_2 b_3-b_3^2-b_1^2 b_4 > 0 $  \\ 
 $ \implies b_1 b_2 b_3 > b_3^2+b_1^2 b_4.$ 
 
For the full expansion of equation (\ref{ctritera1}) for all $\rho_1 > 0, \rho_2  > 0, \zeta_1   > 0, \zeta_2 > 0$, see supplementary material.
  \item
  \begin{equation}
 \begin{aligned}
 &  \bigg( (\rho_1+\rho_2+\zeta_1+\zeta_2-\text{tr(A)} )(-\zeta_1 \text{det(A)}-\zeta_2 \text{det(A)}+\zeta_1 \zeta_2 a_2+(a_{22} a_{33}-a_{23} a_{32}) \rho_1 \zeta_2 \\
 & +(a_{11} a_{33}-a_{13} a_{31}) \rho_2 \zeta_1 ) - ( \zeta_1 \zeta_2 \text{det(A)})\bigg) \bigg((\rho_1+\rho_2+\zeta_1+\zeta_2-\text{tr(A)})( \rho_1 \rho_2+\rho_1 \zeta_1 \\
 & +\rho_2 \zeta_1+\zeta_1 \zeta_2-\zeta_1 \text{tr(A)}-\zeta_2 \text{tr(A)} -(a_{11}+ a_{33}) \rho_2-(a_{22}+a_{33}) \rho_1+a_2 )  \\
 & (-\text{det(A)}+\zeta_1a_2+\zeta_2a_2+(a_{11} a_{33}-a_{13} a_{31}) \rho_2 \\
 & +(a_{22} a_{33}-a_{23} a_{32}) \rho_1-(a_{22}+a_{33}) \rho_1 \zeta_2  -a_{33} \rho_1 \rho_2-a_{33} \rho_2 \zeta_1-a_{11} \rho_2 \zeta_1-\zeta_1\zeta_2 \text{tr(A)} ) \\
& -(-\text{det(A)}+\zeta_1a_2+\zeta_2a_2+(a_{11} a_{33}-a_{13} a_{31}) \rho_2 +(a_{22} a_{33}-a_{23} a_{32}) \rho_1-(a_{22}+a_{33}) \rho_1 \zeta_2\\ 
 & -a_{33} \rho_1 \rho_2-a_{33} \rho_2 \zeta_1-a_{11} \rho_2 \zeta_1-\zeta_1\zeta_2 \text{tr(A)} ))^2 -(\rho_1+\rho_2+\zeta_1+\zeta_2-\text{tr(A)})^2 \\
 &- ( \rho_1+\rho_2+\zeta_1+\zeta_2-\text{tr(A)})(-\zeta_1 \text{det(A)} -\zeta_2\text{det(A)} +\zeta_1 \zeta_2 a_2 +(a_{22} a_{33}-a_{23} a_{32}) \rho_1 \zeta_2 \\
 & +(a_{11} a_{33}-a_{13} a_{31}) \rho_2 \zeta_1 ) \bigg) \\
 & - ( \zeta_1 \zeta_2 \text{det(A)}
)\bigg((\rho_1+\rho_2+\zeta_1+\zeta_2-\text{tr(A)})( \rho_1 \rho_2+\rho_1 \rho_1 +\rho_2 \zeta_1+\zeta_1 \zeta_2-\zeta_1 \text{tr(A)}-\zeta_2 \text{tr(A)} \\
 &-(a_{11}+ a_{33}) \rho_2-(a_{22}+a_{33}) \rho_1+a_2 )- (-\text{det(A)}+\zeta_1a_2+\zeta_2a_2+(a_{11} a_{33}-a_{13} a_{31}) \rho_2 \\
 & +(a_{22} a_{33}-a_{23} a_{32}) \rho_1-(a_{22}+a_{33}) \rho_1 \zeta_2  -a_{33} \rho_1 \rho_2-a_{33} \rho_2 \zeta_1-a_{11} \rho_2 \zeta_1-\zeta_1\zeta_2 \text{tr(A)} )\bigg)^2 \\
 & -(\rho_1+\rho_2+\zeta_1+\zeta_2-\text{tr(A)})(- \zeta_1 \zeta_2 \text{det(A)})^2 > 0
 \end{aligned}
 \label{ctritera2}
 \end{equation}
 $ = (b_1 b_4-b_5)(b_1 b_2 b_3-b_3^2-b_1^2 b_4)-b_5 (b_1 b_2-b_3)^2-b_1 b_5^2  > 0 $ \\ 
$ \implies (b_1 b_4-b_5)(b_1 b_2 b_3-b_3^2-b_1^2 b_4) > b_5 (b_1 b_2-b_3)^2+b_1 b_5^2 $
\\
 For the full expansion of equation (\ref{ctritera2}) for all $\rho_1 > 0, \rho_2  > 0, \zeta_1   > 0, \zeta_2 > 0,$ see supplementary material.
\end{enumerate}
\end{proof}

\newpage
\textbf{Appendix C: Description of the model without quiescence} \\
In this section we will develop a mathematical model that describes the evolution of single Host- two parasites with constant recruitment rate. The model is given by these set (system) of ordinary differential equations, it is call the system without quiescence.  \\
\begin{equation}
\begin{aligned}
& \frac{dI_1}{dt} = \beta_1 H I_1-d I_1-\gamma_1  I_1 -\nu_1I_1+\epsilon_1\\
& \frac{dI_2}{dt} =  \beta_2 H I_2-d I_2-\gamma_2 I_2-\nu_2I_2+\epsilon_2 \\
& \frac{dH}{dt} =  \Lambda- \beta_1H  I_1- \beta_2H I_2 -dH +\nu_1I_1+\nu_2I_2
\end{aligned}
\label{withoutq}
\end{equation}\\ 
\textbf{Steady State Solution of the System} \\ 
The analysis of the steady state of the the system without quiescence (\ref{withoutq}) follows the same steps as for the system with quiescence.

\textbf{Transition Probabilities}
\begin{table}[h!]
\begin{adjustbox}{width=0.95\textwidth,center}
\begin{tabular}{lll}
\hline
Type & Transition & Rate\\
\hline
birth of healthy host $H$ 
& $(H_t,{I_1}_t,{I_2}_t)\rightarrow({H_t}+1,{I_1}_t,{I_2}_t)$
&$\Lambda  \Delta t  +\tiny {o}\Delta(t)$\\
natural death of $H$
& $(H_t,{I_1}_t,{I_2}_t)\rightarrow({H_t}-1,{I_1}_t,{I_2}_t)$
&$ d H \Delta t  +\tiny {o}\Delta(t) $\\
infection of $H$ by $I_1$
& $(H_t,{I_1}_t,{I_2}_t)\rightarrow({H_t}-1,{I_1}_t+1,{I_2}_t)$
&$\beta_1 H I_1  \Delta t +\tiny {o}\Delta(t)$\\
infection of $H$ by $I_2$
& $(H_t,{I_1}_t,{I_2}_t)\rightarrow({H_t}-1,{I_1}_t,{I_2}_t+1)$
&$\beta_2 H I_2  \Delta t +\tiny {o}\Delta(t)$\\
death of $I_1$
&$(H_t,{I_1}_t,{I_2}_t)\rightarrow({H_t},{I_1}_t-1,{I_2}_t)$
&$(d+\gamma_1)I_1  \Delta t+\tiny {o} \Delta(t)$\\
death of $I_2$
&$(H_t,{I_1}_t,{I_2}_t)\rightarrow({H_t},{I_1}_t,{I_2}_t-1)$
&$(d+\gamma_1)I_2  \Delta t+\tiny {o} \Delta(t)$\\
recovery  $I_1$  \& replacement $H$
& $(H_t,{I_1}_t,{I_2}_t)\rightarrow({H_t}+1,{I_1}_t-1,{I_2}_t)$
&$\nu_1I_1 \Delta t +\tiny {o}\Delta(t)$\\
recovery $I_2$ \& replacement $H$
& $(H_t,{I_1}_t,{I_2}_t)\rightarrow({H_t}+1,{I_1}_t1,{I_2}_t-1)$
&$\nu_2I_2 \Delta t +\tiny {o}\Delta(t)$\\
immigration to $I_1$
&$(H_t,{I_1}_t,{I_2}_t)\rightarrow({H_t},{I_1}_t+1,{I_2}_t)$
&$\epsilon_1   \Delta t  +\tiny {o}\Delta(t) $\\
immigration to $I_2$
&$(H_t,{I_1}_t,{I_2}_t)\rightarrow({H_t},{I_1}_t,{I_2}_t+1)$
&$\epsilon_2   \Delta t  +\tiny {o}\Delta(t) $\\
\end{tabular}
\end{adjustbox}
\caption{Transitions for the quiescence model \ref{withoutq}}\label{quiescIItrans2}
\end{table}
%

\textbf{Master equation} \\
Let  $ p(i,j,k)(t)  =  \text{Prob} \{H(t) = i, I_1(t)= j, I_2(t) = k  \}, $ then \\
 \begin{equation}
\begin{aligned}
 \frac{\mathrm dp_{(i,j,k)}}{\mathrm d t}  =  & \Lambda  p_{(i-1,j,k)}+ d (i+1)p_{(i+1,j,k)} +\beta_1( i+1) (j-1)p_{(i+1,j-1,k)} \\
 & +(d  +\gamma_1)(j+1)p_{(i,j+1,k)}+\beta_2 (i+1)(k-1)p_{(i+1,j,k-1)} +(d+\gamma_2) (k+1)p_{(i,j,k+1)} \\
 &+ \nu_1 (j+1)p_{(i-1,j+1,k)}+\nu_2 (k+1)p_{(i-1,j,k+1)} + \epsilon_1 p_{(i,j-1,k)}+ \epsilon_2 p_{(i,j,k-1)} \\
& -\left[ \Lambda+ di+\beta_1 ij 
+( d+\gamma_1)j+\beta_2 ik+(d+\gamma_2) k+\nu_1 j+\nu_2 k + \epsilon_1 + \epsilon_2 \right] p_{(i,j,k)}
\end{aligned}
\label{Kolmo2}
\end{equation}
This master equation (\ref{Kolmo2}) is then used to work out \textit{Kramers-Moyal expansion} that led to the derivation of the \textit{Fokker-Planck equation} below.

\textbf{Derivation of Fokker-Planck Equation}\\
Now, let $$p(i,j,k) =  \int_{ih-\frac{h}{2}}^{ih+\frac{h}{2}}\int_{jh-\frac{h}{2}}^{jh+\frac{h}{2}}\int_{kh-\frac{h}{2}}^{kh+\frac{h}{2}} u(x,y,z)dxdydz+o(h^4),$$
let also $x = ih,y = jh, z= kh$ and $h = \frac{1}{N}$. We then performed \textit{Kramers-Moyal expansion} to derived the following \text{Fokker-Planck equation} which is given as follows.
\begin{equation}
\begin{aligned}
\partial_t u(x,y,t) = -& \partial_x \{h\lambda-dx-\beta_1 xy-\beta_2 xz+\nu_1 y+\nu_2 z\} u(x,y,z) \\
- & \partial_y \{\beta_1 xy-(d+\gamma_1)y-\nu_1 y+h\epsilon_1\} u(x,y,z) \\
 - & \partial_z \{\beta_2 xy-(d+\gamma_2)y-\nu_2 y+ h \epsilon_2\} u(x,y,z) \\
+ &\frac{h}{2} \partial_{xx} \{\lambda+dx+\beta_1 xy+\beta_2 xz+\nu_1 y+\nu_2 z\} u(x,y,z) \\
 - & h \partial_{xy} \{\beta_1 xy+\nu_1 y \} u(x,y,z) \\
+ &\frac{h}{2} \partial_{yy} \{\beta_1 xy+(d+\gamma_1)y+ \nu_1 y+\epsilon_1\} u(x,y,z) \\
 - &h  \partial_{xz} \{\beta_2 xz+\nu_2 z \} u(x,y,z) \\
 + & \frac{h}{2} \partial_{zz} \{\beta_2 xy+(d+\gamma_1)y+ \nu_2 y+\epsilon_2\} u(x,y,z) 
\end{aligned}
\end{equation}
\textbf{Linear Transformation of the Fokker-Planck equation}
\newtheorem*{mydef1}{Theorem}
\begin{mydef1}
The linear Fokker-Planck equation for the above non-linear Fokker-Planck can be written more compactly as follows
\begin{equation}
\frac{\partial P(y,t)}{dt} = - \sum_{ij}^3 M_{ij} \frac{\partial}{\partial y_i} y_i P(y,t)+\frac{1}{2} \sum_{ij}^3 N_{ij} \frac{\partial^2}{\partial y_i \partial y_j}P(y,t) \\
\end{equation}
where $y = (x,y,z), N_{ij}$ is symmetric and positive definite, its solution is give as \\
$$P(y,t) = (2\pi)^\frac{1}{2} det (\Sigma)^\frac{1}{2} exp(-\frac{1}{2} y \Sigma^{-1} y^T) $$
with $$\Sigma^{-1}  = 2\int_0^\infty e^{-Mt} N e^{-Mt} dt. $$
\end{mydef1}
\newtheorem*{mydef2}{Theorem}
\begin{mydef2}
For every matrix $N$ which is symmetric and positive-definite, there a unique solution $\Sigma^{-1}$ to the following equation known as Lyapunov equation $$M \Sigma^{-1}+\Sigma^{-1} M^T  = N$$  where $\Sigma^{-1}$ is symmetric, positive-definite and equal to 
$$\Sigma^{-1}  = \int_0^\infty e^{-Mt} N e^{-M^Tt} dt. $$
\end{mydef2}

The above theorem known as Lyapunov theorem (see [4]) gives us the opportunity to compute covariance matrix more easily since matrices $M$ and $N$ are constant matrices,the only unknown is $\Sigma^{-1}$ matrix.  We use MATLAB to obtain covariance matrix  $\Sigma^{-1}$ numerically. The stochastic matrices $M$ and $N$ for the system without matrix is the same to that of the system with quiescence with the quiescence phase deleted. \\

\newpage
\textbf{ Appendix D: Stochastic Matrices of the Linear Fokker-Planck equation}
\footnotesize{\[
M  = \left(
    \begin{array}{ccccc}
 -d-\beta_1 I_1^*-\beta_1 I_2^*   & -\beta_1 H^*+\nu_1 & -\beta_1 H^* +\nu_2 & 0 & 0  \\ 
  \beta_1 I_1^* &  \beta_1 H^*-d-\gamma_1-\nu_1-\rho_1 & 0 & \zeta_1 & 0\\
 \beta_1 I_2^* &  0 & \beta_1 H^*-d-\gamma_2-\nu_2-\rho_2& 0&\zeta_2 \\ 
 0 & \rho_1 & 0  & -\zeta_1-d & 0 \\ 
 0 & 0 & \rho_2 & 0 &  -\zeta_2-d \\ 
    \end{array}
  \right)
\]
}

\newcommand\scalemath[2]{\scalebox{#1}{\mbox{\ensuremath{\displaystyle #2}}}}
\[ 
N = \left( 
 \scalemath{0.5}{
    \begin{array}{ccccc}
 \lambda+dH^*+\beta_1 H^*I_1^*+\beta_1 H^*I_2+\nu_1 I_1^*+\nu_2 I_2^*  & -(\beta_1 H^*I_1^*+\nu_1 I_1^*) & -(\beta_1 H^*I_2^*+\nu_1 I_2^*) & 0 & 0  \\ 
 -(\beta_1 H^*I_1^*+\nu_1 I_1^*) & \beta_1 H^*I_1^*+(d+\gamma_1)I_1^*+\nu_1 I_1^*+\rho_1 I_2^*+\zeta_1 Q_1^*+\epsilon_{11} & 0  & -(\rho_1 I_1^*+\zeta_1 Q_1^*) & 0\\
-(\beta_1 H^*I_2^*+\nu_1 I_2^*) &  0 & \beta_1 H^*I_2^*+(d+\gamma_2)I_2^*+\nu_2 I_2^*+\rho_2 I_2^*+ \zeta_2 Q_2^* + \epsilon_{12} & 0&-(\rho_2 I_2^*+\zeta_2 Q_2^*) \\ 
 0 &-( \rho_1 I_1^*+\zeta_1 Q_1^*) & 0  & \rho_1 I_1^*+\zeta_1 Q_1^*+dQ_1^* & 0 \\ 
 0 & 0 & -(\rho_2 I_2^*+\zeta_2 Q_2^*) & 0 &  \rho_2 I_2^*+\zeta_2 Q_2^*+dQ_2^* 
    \end{array}}
  \right)
\]
where $H^*,I_1^*,I_2^*,Q_1^*,Q_2^*$ are equilibrium solutions of \ref{Model1} (rearranged in such away that healthy compartment comes first equation in the system. The order of the other compartments remains unchanged).

\end{document}